\documentclass[aps,pra,amsmath,superscriptaddress,reprint,floatfix,notitlepage,balancelastpage,twocolumn,reprint]{revtex4}
\usepackage[colorlinks,urlcolor=red,citecolor=red]{hyperref}
\usepackage{graphicx,subfigure} 
\usepackage{epstopdf}
\usepackage{amsthm} 

\usepackage{tikz}

\let\baraccent=\= 
\renewcommand{\=}[1]{\stackrel{#1}{=}} 

\theoremstyle{definition}

\theoremstyle{remark}

\newcommand{\Deltat}{\tilde{\Delta}}

\newcommand{\qt}{\tilde{q}}

\newcommand{\qh}{\hat{q}}\newcommand{\Sh}{\hat{S}}
\newcommand{\qb}{\bar{q}}

\newcommand{\nf}{n_{\rm F}}

\newcommand{\tc}{{T_{\rm c}}}

\newcommand{\phdag}{{\phantom{\dagger}}}

\newcommand{\ef}{\epsilon_{\rm F}}

\newcommand{\curO}{{\cal O}}

\newcommand{\kb}{k_{\rm B}}

\newcommand{\bq}{{\bf q}}
\newcommand{\bp}{{\bf p}}

\newcommand{\ph}{\hat{p}}

\newcommand{\xh}{\hat{x}}

\newcommand{\hfflo}{h_{\rm FFLO}}
\newcommand{\pfflo}{P_{\rm FFLO}}

\newcommand{\br}{{\bf r}}

\newcommand{\be}{\begin{equation}}
\newcommand{\ee}{\end{equation}}
\newcommand{\bea}{\begin{eqnarray}}
\newcommand{\eea}{\end{eqnarray}}
\newcommand{\bse}{\begin{subequations}}
\newcommand{\ese}{\end{subequations}}

\newcommand{\txi}{\tilde{\xi}}
\newcommand{\mut}{\tilde{\mu}}

\begin{document}

\title{Fulde-Ferrell-Larkin-Ovchinnikov state of two-dimensional imbalanced Fermi gases}
\author{Daniel E. Sheehy}
\email{sheehy@lsu.edu}
\affiliation{Department of Physics and Astronomy, Louisiana State University, Baton Rouge, LA, 70803, USA}
\date{July 16, 2014}
\begin{abstract} The ground-state phase diagram of attractively-interacting Fermi gases in two dimensions with 
a population imbalance is investigated.  We find the regime of stability for the 
Fulde-Ferrell-Larkin-Ovchinnikov (FFLO) phase, in which pairing  occurs at finite wavevector,
and determine the magnitude of the pairing amplitude $\Delta$ and FFLO wavevector $q$ in the ordered
phase, finding that $\Delta$ can be of the order of the two-body binding energy.
  Our results rely on a careful analysis of the zero temperature gap equation for the FFLO state, 
which possesses nonanalyticities as a function of $\Delta$ and $q$, invalidating a Ginzburg-Landau 
 expansion in small $\Delta$.
\end{abstract}


\maketitle

\section{Introduction}
The Fulde-Ferrell-Larkin-Ovchinnikov (FFLO) state is a superfluid phase that can occur
when two species (or spin-state $\sigma = \uparrow, \downarrow$) of fermion pair and condense in the presence of a density imbalance~\cite{FF,LO}
that is parameterized by the polarization $P=(N_\uparrow- N_\downarrow)/(N_\uparrow- N_\downarrow)$ with  $N_\sigma$ the number of 
species $\sigma$.
The FFLO phase has been predicted to occur in a 
wide range of systems, including electronic materials and high-density quark matter~\cite{Casalbuoni,Alford}.
While recent experiments report evidence of the FFLO state in organic materials~\cite{Mayaffre} and in 
thin-film electronic superconductors~\cite{Prestigiacomo},
definitive signatures of the FFLO state (such as the predicted spatially-varying pairing amplitude)
remain elusive. 

Experimental advances in atomic physics have recently led to another
setting for observing the FFLO state, namely cold atomic
gases~\cite{Bloch,Giorgini}, which exhibit several experimentally tunable parameters,
including the fermion densities, the
interfermion interactions, and the effective spatial dimension of the
gas (controlled by an applied  trapping potential).  

However, cold atom experiments in the three-dimensional (3D) limit observed no signatures of the FFLO 
phase~\cite{Zwierlein2006,Partridge2006,Shin2006,Partridge2006prl},
consistent with theoretical work that found the FFLO to be stable for a very narrow range of density imbalance~\cite{SRPRL,Parish2007,SRAOP}.
Imbalanced gases in 1D, studied experimentally at Rice~\cite{LiaoRittner}, are predicted to have a wide parameter
region of stable imbalanced superfluidity~\cite{Orso07,HuLiuDrummond,Feiguin,LiuPRA2007,Batrouni,HM2010,SunBolech}, although 
 finite-momentum pairing correlations (signifying the FFLO state) have not been directly observed.

\begin{figure}[ht!]
     \begin{center}
        \subfigure{
            \includegraphics[width=85mm]{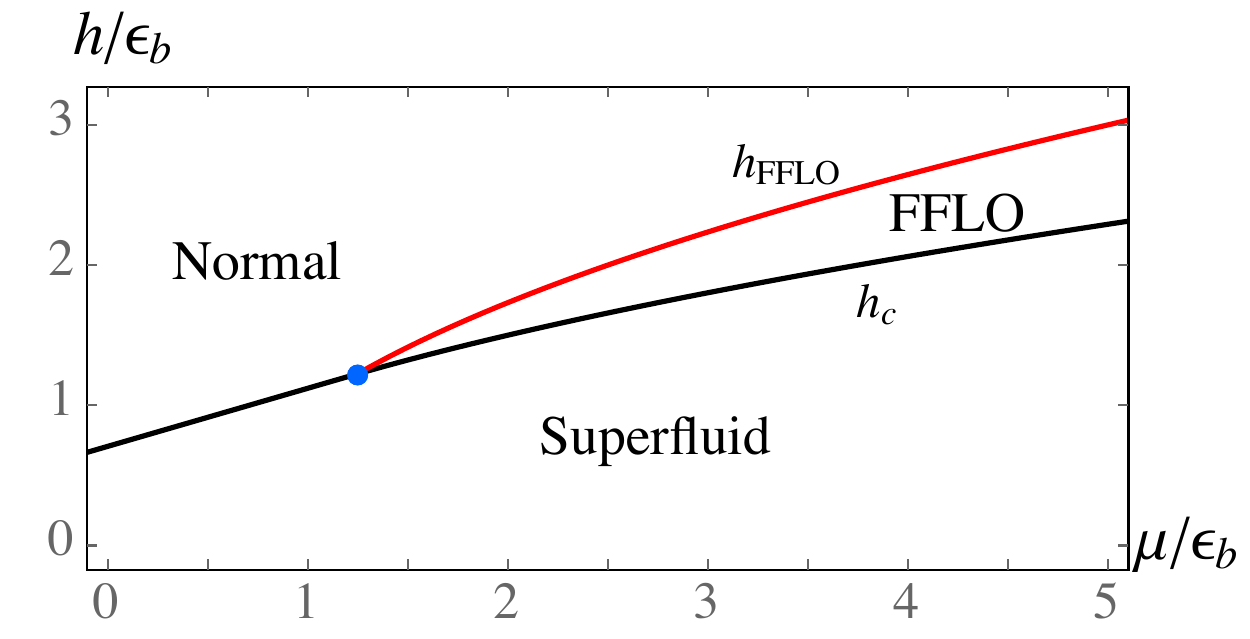}}
        \\ \vspace{-.5cm}
        \subfigure{
            \includegraphics[width=85mm]{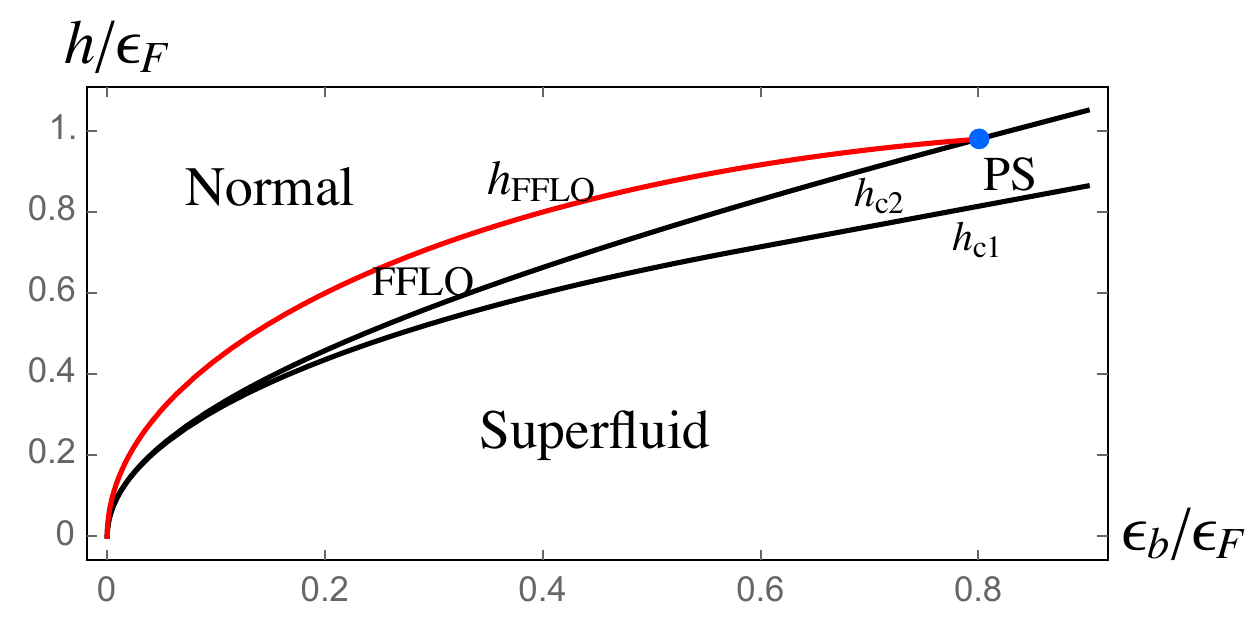}}
    \end{center}\vspace{-.5cm}
    \caption{
        (Color Online) The top panel shows the phase diagram for 2D imbalanced Fermi gases at fixed total chemical potential
$\mu$ and Zeeman field $h$ (normalized to the two-body binding energy $\epsilon_b$). The bottom panel shows the phase diagram for
fixed total particle number $N=N_\uparrow+N_\downarrow$, as a function of binding energy $\epsilon_b$ normalized to the Fermi energy.
     }
\vspace{-.5cm}
   \label{fig:one}
\end{figure}

In this paper we investigate Fermi gases in two spatial dimensions, which, in the balanced case, have been studied theoretically in
Refs.~\cite{Randeria89,Petrov2003} and experimentally in Refs.~\cite{Sommer2012,Zhang2012,Ries,Murthy}.   Imbalanced 2D fermion superfluids have been
studied theoretically for many years in the condensed matter context~\cite{Shimahara,Mora,Loh,Aoyama} and, more recently, by several authors in
the present cold-atom context~\cite{Conduit,HeZhuang,RV,Leo2011,Wolak2012,LevinsenBaur,Caldas,ParishLevinsen,Yin}.  
Here, our main interest is in determining the phase diagram of 2D Fermi gases as a function of the population imbalance (that is controlled in
cold-atom experiments) and the chemical potential difference $h = \frac{1}{2}(\mu_\uparrow - \mu_\downarrow)$
 (that is controlled in condensed matter experiments via the Zeeman effect).   In addition, we investigate the onset of pairing  in the 
FFLO phase of imbalanced 2D gases.  Interestingly, it is found that 
the equation controlling the pairing amplitude  $\Delta$ in the FFLO phase possesses
a nonanalytic dependence on $\Delta$ that precludes a simple Taylor expansion of the ground-state energy in small $\Delta$.
 Before presenting our detailed calculations, in the next section we first briefly describe our main results.

\subsection{Summary of Main Results}

Using a model Hamiltonian for two species (labeled by $\uparrow,\downarrow$) of fermions in 2D
with  chemical potentials  $\mu_\uparrow = \mu+h$ and $\mu_\downarrow = \mu -h$,
we find the ground-state phase diagram (Fig.~\ref{fig:one} top panel) as a function 
of $\mu/\epsilon_b$ and $h/\epsilon_b$, where $\epsilon_b$ is the
two-body binding energy characterizing the tunable interactions.  As seen in Fig.~\ref{fig:one} (top panel), we find
a window of FFLO stability that widens with increasing $h$,
 between a balanced~\cite{Balancednote} superfluid (SF) and an imbalanced nonsuperfluid or normal (N) phase.
The red curve denotes a continuous N-FFLO transition, and the black curve
denotes a first order N-SF transition for $\mu/\epsilon_b<5/4$ and a first order FFLO-SF transition for $\mu/\epsilon_b>5/4$.
%

\begin{figure}[ht!]
     \begin{center}
 \includegraphics[width=85mm]{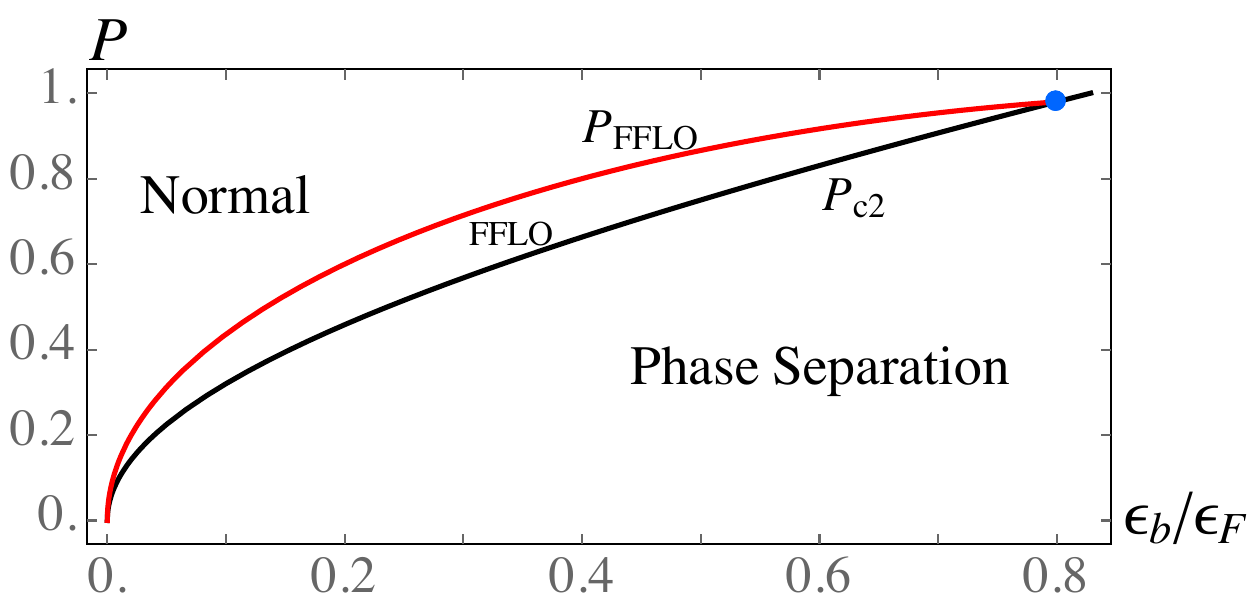}
%
    \end{center}\vspace{-.5cm}
    \caption{
        (Color Online) Phase diagram at fixed total particle number and fixed polarization $P$, as a function of binding energy
normalized to the Fermi energy.
     }
\vspace{-.25cm}
   \label{fig:PCphase}
\end{figure}

While the top panel of Fig.~\ref{fig:one} shows the phase diagram in the fixed chemical potential (grand canonical)
ensemble, the bottom panel shows the phase diagram at fixed total density and fixed chemical potential difference $h$.  This
case is appropriate for describing electronic superconductors, in which $h$
arises from the Zeeman coupling between an external magnetic field and the electron spin.  The axes in Fig.~\ref{fig:one}
are now normalized to the Fermi energy (defined below) that is related to the imposed particle number.  We see that,
by fixing $N$, the first order transition becomes a regime of phase separation and that the regime of stability for
the FFLO phase is confined to small $\epsilon_b<\frac{4}{5}\ef$ (consistent with the 3D case, where the FFLO is only
stable in the weakly-interacting BCS regime~\cite{SRPRL,Parish2007,SRAOP}). Fig.~\ref{fig:PCphase} shows the $T=0$ phase diagram at fixed total 
particle number and fixed polarization $P$, the case 
that is appropriate for cold atom experiments at fixed $N_\uparrow$ and $N_\downarrow$. (Although we do not take into account the harmonic
trapping potential that is typically present in such experiments.)  Since the SF state is unpolarized, the
entire SF region of the phase diagram in Fig.~\ref{fig:PCphase} is confined to $P=0$.

\begin{figure}[ht!]
     \begin{center}
        \subfigure{
            \includegraphics[width=85mm]{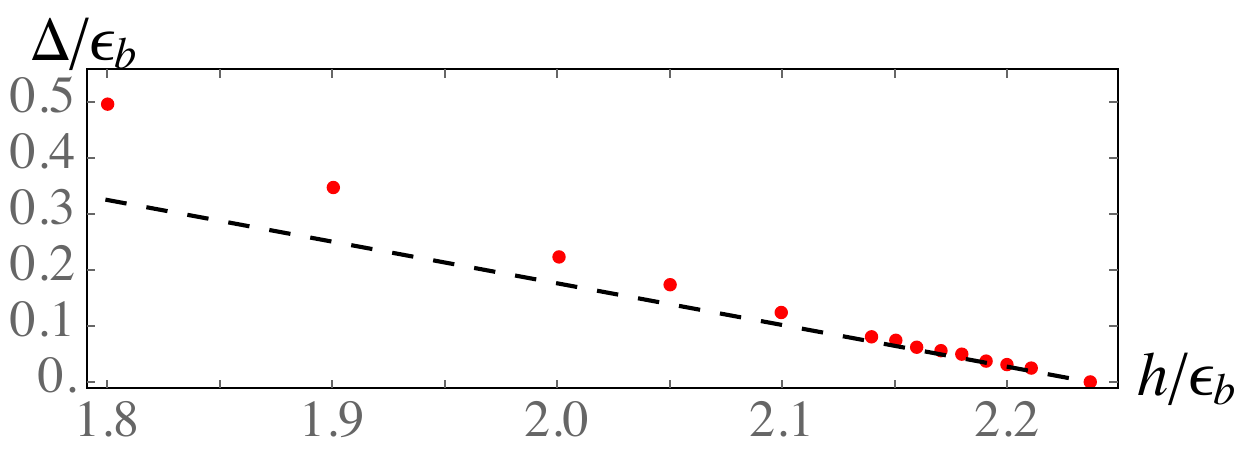}}
        \\ \vspace{-.45cm}
        \subfigure{
            \includegraphics[width=85mm]{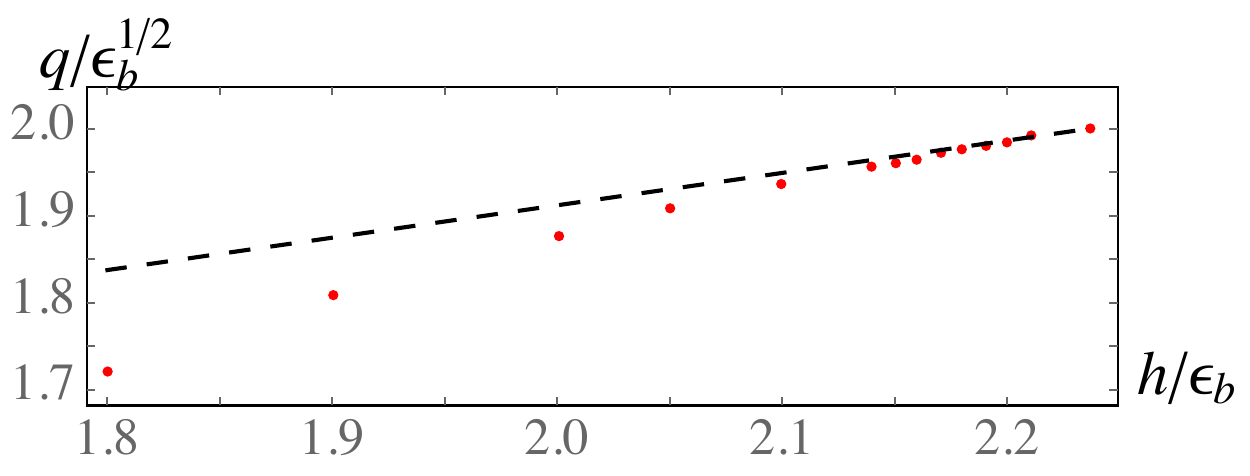}}
    \end{center}\vspace{-.5cm}
    \caption{ (Color Online)  FFLO pairing amplitude $\Delta$ (top
panel) and wavevector $q$ (bottom panel) for the case of $\mu =
3\epsilon_b$.  For this case, $\hfflo = \sqrt{5}\epsilon_b \simeq 2.24\epsilon_b $
and $h_c\simeq 1.8\epsilon_b$. The red dots come from a direct numerical
minimization of Eq.~(\ref{eq:eggeneral}), except for the points
($\Delta = 0$, $q = 2\sqrt{\epsilon_b}$) at $h=\hfflo$ following from
the continuous nature of the transition.  The dashed lines are the
approximate formulas Eq.~(\ref{eq:subresults}).  
} 
\vspace{-.25cm}
   \label{fig:two}
\end{figure}

The FFLO state occuring for $h$ below $\hfflo$ is assumed to be of the single plane-wave form, $\Delta(\br) = \Delta{\rm e}^{i\bq\cdot \br}$, 
characterized by the pairing amplitude $\Delta$  and the FFLO wavevector $q$.   In Fig.~\ref{fig:two} we show our results for
$\Delta$ and $q$ below $\hfflo$, with the dashed lines being approximate analytical formulas for these quantities (valid asymptotically close to the transition) and the red dots following from a numerical minimization of the mean-field free energy.  

As we will show, the value of the FFLO wavevector at the transition, $q_c=2\sqrt{\epsilon_b}$.  This result was also found in Ref.~\cite{Conduit}.  Additionally, the phase diagrams shown in 
Figs.~\ref{fig:one} and ~\ref{fig:PCphase} are consistent with earlier work~\cite{Shimahara,Conduit,Mora,HeZhuang,Caldas}.  Here, we show that the FFLO phase boundary, and the universal
FFLO wavevector, follow from a nonanalytic structure of the integral governing the FFLO gap equation. 
We present a detailed
analysis of the structure of the gap equation for the FFLO phase which possesses kinks (slope discontinuities) as a function of
$\Delta$ (see Fig.~\ref{fig:ess} below). To understand this behavior, we use the Jensen formula from complex analysis, which 
relates integrals on the unit circle in the complex plane to zeroes of the integrand.  
  Despite the nonanalytic structure of
the function governing $\Delta$, we find a continuous transition into the FFLO state, as seen in Fig.~\ref{fig:two}, with
the onset of $\Delta$ and the deviation of $q$ from $q_c$ being linear in $|h-\hfflo|$.

\section{Model Hamiltonian}
We now proceed to derive these results, starting with the Hamiltonian
\bea
&&H =  \sum_{\sigma=\uparrow,\downarrow}\int d^2 r \Psi^\dagger_\sigma(\br) \big(\frac{\ph^2}{2m}-\mu_\sigma\big)
\Psi_\sigma(\br) 
\nonumber 
\\
&&
+ \lambda \int d^2 r \, \Psi^\dagger_\uparrow(\br) \Psi^\dagger_\downarrow(\br)  \Psi_\downarrow^\phdag(\br) \Psi_\uparrow^\phdag(\br),
\eea
where $\Psi_\sigma(\br)$ are field operators for fermions of spin $\sigma$ 
and  $\lambda$ parameterizes the short-range attractive interactions.
%
%
To describe the phases of an imbalanced Fermi gas in 2D, we proceed by making the standard mean-field approximation,
assuming an expectation value $\langle \Psi_\downarrow(\br) \Psi_\uparrow(\br)\rangle = \Delta {\rm e}^{i\bq\cdot \br}$, with
$\Delta$ and $\bq$ being variational parameters that must be minimized, at fixed $\mu_\uparrow$ and $\mu_\downarrow$,
to determine the equilibrium state.  The resulting mean-field free energy is: 
\be
F=
-  \frac{|\Delta|^2}{\lambda} 
- \kb T \sum_{\bp,\alpha = \pm} \ln \big( 1+ {\rm e}^{-\beta E_{\bp\alpha}}\big)
- \sum_\bp \big(E_p - \txi_p\big),
\label{Eq:mffree}
\ee
with $\beta = \frac{1}{\kb T}$.  Here, we defined
\bea
\label{eq:ekpm}
E_{\bp \pm} &=& E_p \pm\big( h + \frac{\bp \cdot \bq}{2m}\big) ,
\\
E_p &=& \sqrt{\txi_p^2 + |\Delta|^2},\label{eq:eepee}
\\
\txi_p &=& \epsilon_p - \mut,
\eea
where 
  $\epsilon_p = \frac{p^2}{2m}$ and 
$\mut = \mu-q^2/8m$.
Below we often take $m=1$, and $\Delta$ real and positive.
  In the zero-temperature limit, this becomes the ground-state energy~\cite{SRAOP}
\be
\label{eq:eggeneral}
E_G = - \frac{|\Delta|^2}{\lambda} - \sum_\bp \big( E_p - \txi_p) + \sum_{\bp, \alpha = \pm} 
E_{\bp\alpha}\Theta(-E_{\bp\alpha}),
\ee
where $\Theta(x)$ is the Heaviside step function.

 Our model Hamiltonian must include a ultraviolet scale $D$ cutting off all momentum sums.
As is standard~\cite{Randeria89}, we express $\lambda$ in terms of the two-body binding energy $\epsilon_b$,
satisfying
\be
\label{eq:relation}
\frac{1}{\lambda} =\sum_\bp \frac{1}{2\epsilon_p+ \epsilon_b},
\ee
where the sum is over $\epsilon_p<D$, leading to $\epsilon_b \simeq 2D\exp[m/\pi\lambda]$.
Upon inserting Eq.~(\ref{eq:relation}) into Eqs.~(\ref{Eq:mffree}) and (\ref{eq:eggeneral}), we can 
take $D\to \infty$ in the momentum sums, with all dependence on $D$ absorbed into $\epsilon_b$.

\subsection{Gap equation}
To find the phase diagram we need to minimize $F$ with respect to the
variational parameters $q$ and $\Delta$.  The minimization with respect to $\Delta$ defines a function $S(h,q,\Delta)$ via
$\frac{\partial F}{\partial \Delta}=-2\Delta S(h,q,\Delta)$ that is given by 
\bea
\nonumber
&& S(h,q,\Delta) = 
\int \frac{d^2 p}{(2\pi)^2} \Big[\frac{1-\nf(E_{\bp+}) - \nf(E_{\bp-})}{E_{\bp+}+E_{\bp-}} 
\\
&& \qquad \qquad \qquad 
- \frac{1}{2\epsilon_p+|\epsilon_b|}\Big],
\label{Eq:fulldelta}
\eea
where we used Eq.~(\ref{eq:relation}).  
Here, $\nf(x) = \frac{1}{{\rm e}^{x/T}+1}$ is the Fermi function.
Equilibrium FFLO superfluid states satisfy $S(h,q,\Delta)=0$ (also known as the 
gap equation) as well as a 
similar equation coming from minimizing with respect to $q$ ($\frac{\partial F}{\partial q}=0$).  Our main focus will
be on evaluating $S(h,q,\Delta)$ which, as we'll show, has a nonanalytic dependence on $\Delta$ in the $T\to 0$ limit.

As we will show, in the parameter region of interest (where the FFLO is stable), $S(h,q,\Delta)$ 
is {\em independent\/} of $\Delta$ for small $\Delta$, exhibiting kinks (where its slope is discontinuous) at two values of 
$\Delta$.  This implies that a simple Ginzburg-Landau type expansion, based on a Taylor expansion in small
$\Delta^2$:
\be
\label{eq:tayloress}
S(h,q,\Delta) \simeq S(h,q,0) +\Delta^2 S'(h,q,0),
\ee
with $S'(h,q,0) = \frac{\partial S}{\partial \Delta^2} \Big|_{\Delta^2 \to 0}$, will fail.  Although these nonanalyticities are
smoothed out by finite temperature, at low $T$ this function is  \lq\lq almost\rq\rq\ nonanalytic, in the sense that Eq.~(\ref{eq:tayloress})
yields a poor approximation to the  gap equation integral $S(h,q,\Delta)$.
%
%

%

\section{Phase diagram at $q=0$}

In this section, we determine the phase boundaries assuming $q=0$ pairing, which 
amounts to neglecting the possibility of the FFLO state.   
 Within this approximation, 
Eq.~(\ref{eq:eggeneral}) exhibits, with increasing $h$, a first-order transition from a fully-paired
balanced phase to an imbalanced normal phase, labeled $h_c$ in Fig~\ref{fig:one} (top panel).  We will
also take the zero temperature limit.

\subsection{Balanced paired phase}
We first recall the balanced paired phase~\cite{Randeria89} by
setting $h=0$, yielding the gap equation $0= -2\Delta S(0,0,\Delta)$ with 
%
(converting the momentum sum to an integration, with the system area set to unity)
\be
S(0,0,\Delta) = \int \frac{d^2 p}{(2\pi)^2} \Big[\frac{1}{2\sqrt{\xi_p^2 + \Delta^2}}- \frac{1}{2\epsilon_p + 
\epsilon_b}\Big].
\label{eq:essder}
\ee
Thus, in this limit both Fermi functions in Eq.~(\ref{Eq:fulldelta}) vanish.
Evaluation of the integral in Eq.~(\ref{eq:essder}) leads to  the stationary pairing amplitude~\cite{Randeria89} (a minimum of
$E_G$) at
\be 
\label{Eq:statgap}
\Delta = \sqrt{\epsilon_b}\sqrt{2\mu+\epsilon_b},
\ee
so that we must have $\mu>-\frac{1}{2}\epsilon_b$ for a stable SF phase.  The total particle number of
the SF state at fixed $\mu$ is given by  computing $N = - \frac{\partial E_G}{\partial \mu}$, 
yielding for the total density
\be
n = \frac{1}{2\pi} \big(\mu+\sqrt{\Delta^2+\mu^2}\big).
\ee
  Combining the preceding expressions yields results for $\Delta$ and $\mu$ for a system at fixed imposed 
particle number:
\bse
\label{Eq:crossover}
\bea
\frac{\Delta}{\ef} &=& \sqrt{2\frac{\epsilon_b}{\ef}} ,
\\
\frac{\mu}{\ef} &=& 1 - \frac{1}{4} \big(
\frac{\Delta}{\ef} 
\big)^2 = 1- \frac{1}{2} \frac{\epsilon_b}{\ef},
\label{Eq:crossover2}
\eea
\ese
where the Fermi energy $\ef = \pi n$ is  defined by the value of $\mu$ in the normal phase ($\Delta = 0$).

\subsection{Imbalanced case at fixed chemical potential}
Having briefly reviewed the balanced case, we now consider the imbalanced case $h>0$ (but still with $q=0$). 
We find that the
location of the minimum is unchanged with increasing $h$, although a second minimum of $E_G$,
at $\Delta=0$, appears describing the imbalanced normal phase.  The location, $h_c$, of the first order
transition between the SF and N phases is obtained by equating the energies, $E_{G,SF}=E_{G,N}$.  To find
$E_{G,SF}$, we evaluate the momentum sum in Eq.~(\ref{eq:eggeneral}) (note the final term vanishes in this phase) and
 insert the stationary gap-equation solution, obtaining (note $m=1$)
\be
E_{G,SF} = - \frac{1}{2\pi} \big(\mu+\frac{1}{2}\epsilon_b\big)^2,
\ee
where we assumed $\mu>-\frac{1}{2}\epsilon_b$.  To find $E_{G,N}$, we set  $\Delta = 0$ in Eq.~(\ref{eq:eggeneral})
and evaluate the momentum sums. The result is 
\be
E_{G,N} = - \frac{1}{4\pi} \Big[(\mu+h)^2 \Theta(\mu+h) + (\mu-h)^2 \Theta(\mu-h)\Big].
\ee
Since we always assume $\mu+h>0$, the first term in square brackets is nonzero.  The second term is 
zero if the imbalanced normal phase has only one Fermi surface (i.e., it is a single-species Fermi gas)
and nonzero if the imbalanced
normal phase has two Fermi surfaces (i.e., it is a two-species Fermi gas).  Our result for $h_c(\mu)$ is
different depending on whether $\mu+h>0$ at the transition.  Equating the SF and imbalanced normal
energies yields
\be
h_c(\mu) = \begin{cases}\epsilon_b \big[\frac{1}{\sqrt{2}} + 
\big( \sqrt{2}-1  \big)\mu/\epsilon_b\big] &\text{for} \,\, \mu<\mu_c, \,\, \cr
\frac{1}{2} \epsilon_b \sqrt{1+ 4\mu/\epsilon_b}&\text{for} \,\, \mu>\mu_c, \,\,
\end{cases} 
\label{eq:hcmu}
\ee
%
where  $\mu_c=\frac{1}{2}(1+\sqrt{2})\epsilon_b$ separates the cases of a transition into a phase with 
one (for $\mu<\mu_c$) or two (for $\mu>\mu_c$) Fermi surfaces.  This determines the black curve of Fig.~\ref{fig:one} (top panel).
As discussed above, for $\mu/\epsilon_b>5/4$, the actual first order transition is between the FFLO and SF phases.  However, we find
(numerically) that the FFLO and N phases are almost equal in energy, so that the true FFLO-N first order phase boundary is only slightly
lower than Eq.~(\ref{eq:hcmu}).

\subsection{Imbalanced case at fixed particle number}
Our next task is to consider the case of fixed particle number.  There are two cases to consider, fixed total particle number and
fixed Zeeman field $h$ (appropriate for a electronic thin-film superconductor in an in-plane magnetic field) and fixed total particle
number and fixed magnetization (or fixed polarization, appropriate for cold-atom realizations).  We start with the first case.

Imposing fixed total particle number causes the first order phase boundary in the fixed-$\mu$ ensemble to
open up into a region of phase separation bounded by curves $h_{c1}<h_{c2}$ as shown in Fig.~\ref{fig:one} (bottom panel).  
The lower critical Zeeman field $h_{c1}$ is defined
by approaching the phase transition from the SF regime at fixed particle number.  Inserting the SF phase chemical potential 
Eq.~(\ref{Eq:crossover2}) into Eq.~(\ref{eq:hcmu}) leads to:
\be
\frac{h_{c1}}{\ef} = \begin{cases}
\frac{1}{2} \frac{\epsilon_b}{\ef}  + \sqrt{2} - 1
  & \text{for} \,\,
\frac{\epsilon_b}{\ef} >\frac{\sqrt{2}}{\sqrt{2}+1},
\cr
 \frac{1}{2} \frac{\epsilon_b}{\ef}\sqrt{4\frac{\ef}{\epsilon_b} -1}
 &\text{for} \,\, \frac{\epsilon_b}{\ef} <\frac{\sqrt{2}}{\sqrt{2}+1}.
\end{cases}
\label{Eq:hcone}
\ee
The upper critical Zeeman field $h_{c1}$ is similarly defined by approaching $h_c$ assuming the chemical potential is given by 
its value in the imbalanced normal phase.  By solving the particle number equation, we find that
the system chemical potential $\mu = \ef$ in the imbalanced normal phase.  This finally leads to
\be
\frac{h_{c2}}{\ef} = \begin{cases}
 \frac{\varepsilon_b}{\ef} \Big[
\frac{1}{\sqrt{2}} + \big( \sqrt{2}-1  \big)\frac{\ef}{\varepsilon_b} 
\Big]
  & \text{for} \,\,
\frac{\varepsilon_b}{\ef} >\frac{2}{\sqrt{2}+1},
\cr
 \frac{1}{2} \frac{\varepsilon_b}{\ef} \sqrt{1+ 4\frac{\ef}{\varepsilon_b} }
 &\text{for} \,\, \frac{\varepsilon_b}{\ef} <\frac{2}{\sqrt{2}+1}.
\end{cases}
\label{Eq:hctwo}
\ee
for the upper critical Zeeman field at fixed total particle number.

Finally, we turn to deriving the phase boundaries at fixed total particle number and fixed imbalance (or fixed polarization $P$).  
The
phase boundaries $h_{c1}$ and $h_{c2}$ at fixed Zeeman field then become phase boundaries $P_{c1}$ and $P_{c2}$.
  However, $P_{c1} = 0$ since
the SF phase is balanced.  For $P_{c2}$, we need to determine the system polarization in the imbalanced normal 
phase when $h_{c2}$ is reached.
In the regime where the phase transition is into a fully polarized $N$ phase, the first case of Eq.~(\ref{Eq:hctwo}),
 we'll have $P_{c2}=1$.
In the second case of  Eq.~(\ref{Eq:hctwo}), we use that the magnetization $M = \frac{1}{\pi} h$  
in the partially polarized N phase, 
leading to
\be
P_{c2}= \frac{1}{2} \frac{\epsilon_b}{\ef}\sqrt{1+4\frac{\ef}{\epsilon_b}},
\ee
that we plot in Fig.~\ref{fig:PCphase}.  As we have mentioned, this result was calculated assuming a first order transition
between the SF and N while the actual transition is between the SF and FFLO.  However, since the ground state energy
of the FFLO is only slightly lower than the N phase, the actual $P_{c2}$ is only slightly lower than the result displayed
in Fig.~\ref{fig:PCphase}.

Thus, neglecting the FFLO phase, there are only two homogeneous phases in the ground-state phase diagram, the 
imbalanced normal phase and the balanced superfluid phase.  In the 3D case, an additional phase is possible, the magnetic 
superfluid phase~\cite{SRPRL}.  The  $SF_{\rm M}$
 occurs when the Zeeman field $h$ exceeds $\Delta$, leading to a partial depopulation of the pair condensate.
 This ground state is restricted to the deep BEC regime for a 3D gas since, away from this regime, it is preempted
by a first order phase transition.  In the present 2D case, we find that the uniform $SF_{\rm M}$ does not occur
anywhere in the phase diagram.  However, we do find a stable FFLO phase over a wide range of parameters, as we now show.

\section{FFLO phase boundary}

To study the FFLO phase, we first assume a continuous transition, with decreasing $h$, from the $N$ phase
to the FFLO state, occurring when the curvature of $E_G$ vs. $\Delta$, at $\Delta = 0$, becomes zero for
some nonzero $q$. 
This is equivalent to $0= S(h,q,0)$ with
\bea
S(h,q,0)
%
%
=  \int \!\frac{d^2p}{(2\pi)^2}\!\Big[ \frac{N(\bp)}
{\xi_{\bp-\frac{1}{2}\bq \uparrow} + \xi_{\bp+\frac{1}{2}\bq \downarrow} } \!
-\!
\frac{1}{2\epsilon_p+\epsilon_b} \Big],
\label{Eq:simpler}
\eea
where we defined the numerator $N(\bp) = 1-\nf(\xi_{\bp-\frac{1}{2}\bq \uparrow}) 
 - \nf(\xi_{\bp+\frac{1}{2}\bq \downarrow})$ 
(for the moment generalizing to nonzero temperature $T$).  We note that the assumption of a continuous transition will be confirmed below (within our choice
of the single plane-wave FFLO ansatz) in Sec.~\ref{sec:as} where we study the gap equation for $h<\hfflo$ and find $\Delta\to 0$ 
for $h\to \hfflo^-$.
%
%

\begin{figure}[ht!]
     \begin{center}
        \subfigure{
            \includegraphics[width=41mm]{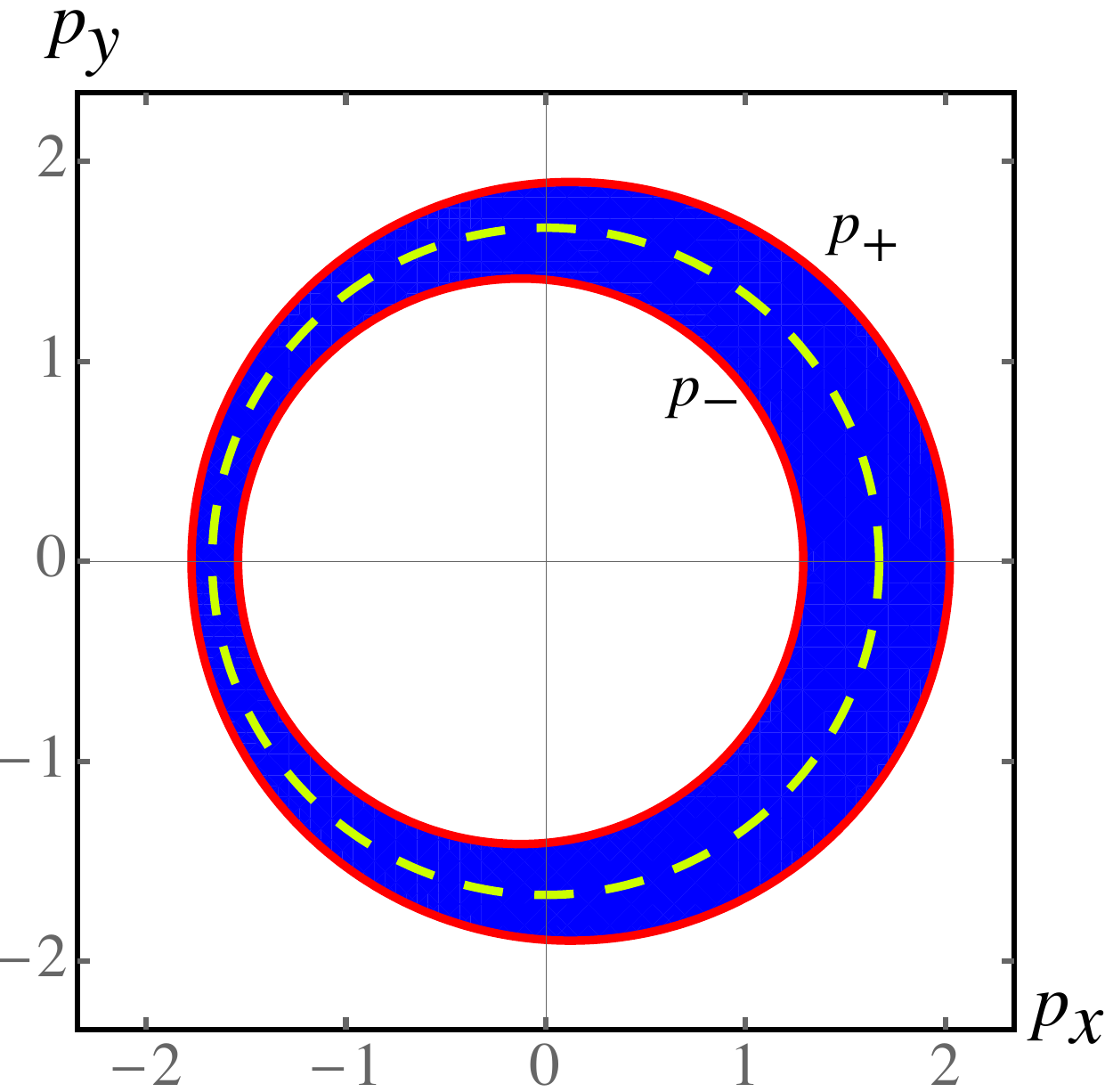}}
        \subfigure{
            \includegraphics[width=41mm]{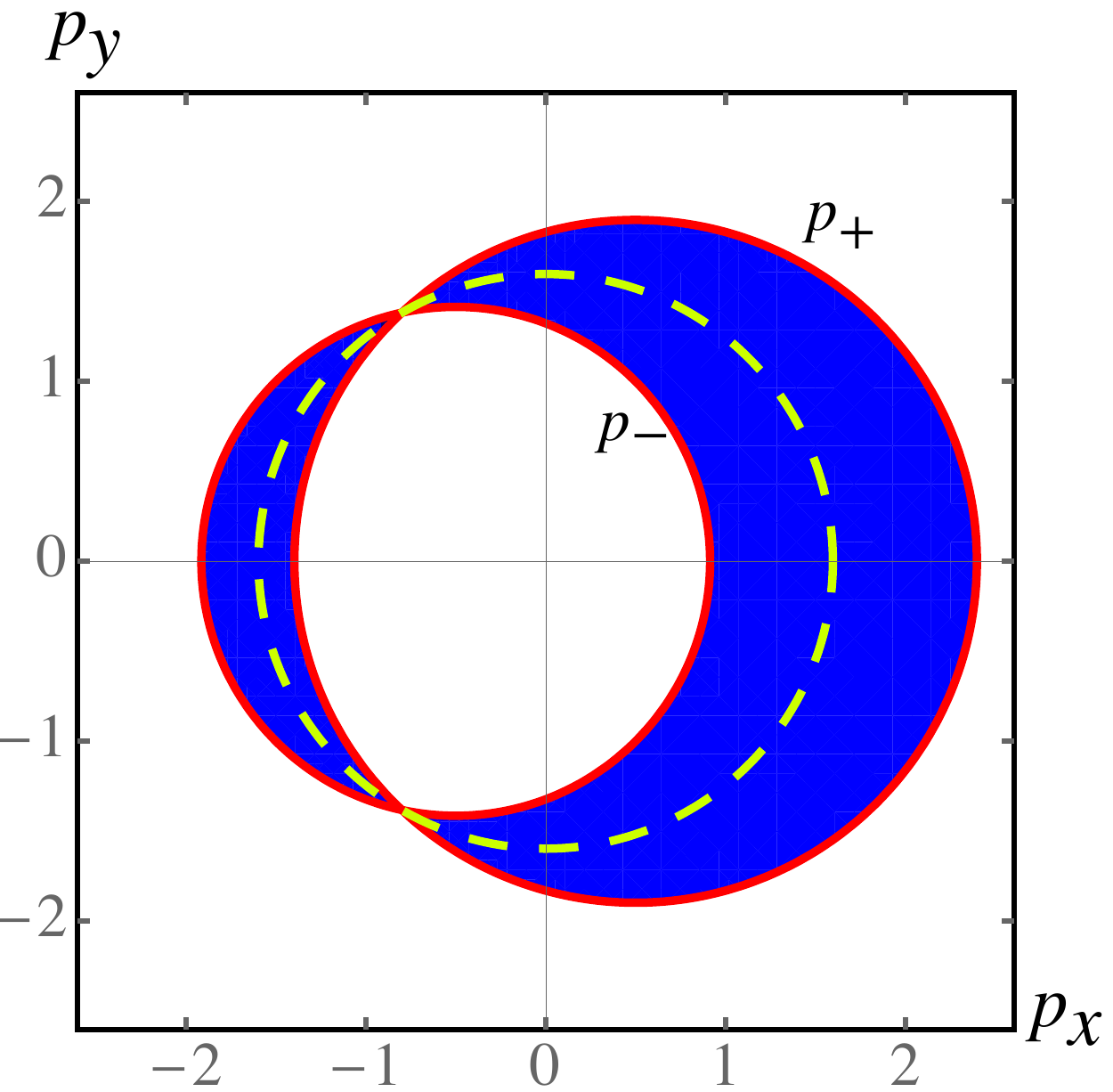}}
    \end{center}\vspace{-.5cm}
    \caption{
        (Color Online) Plots of the curves $p_{\pm}(\theta)$, Eq.~(\ref{Eq:ppm}), for $\mu = 1.4$ and $h=.4$
with $q=0.25$ (so that $q<q_c$) in the left panel and  $q=0.4$ (so that $q>q_c$).  Physically, these correspond
to the Fermi surfaces for the spin-$\uparrow$ and spin-$\downarrow$ fermions, boosted by $\pm q/2$.
  The yellow dashed curve is
$0=\xi_{\bp-\frac{1}{2}\bq \uparrow} + \xi_{\bp+\frac{1}{2}\bq \downarrow}$ (or $p^2=2\mut$), the curve along which the 
denominator of the first term of Eq.~(\ref{Eq:simpler}) vanishes.   
For an intermediate $q$ (at $q_c$), it is clear that the circles will touch at one point; this is the case at the
FFLO phase transition.   
   }\vspace{-.05cm}
   \label{fig:contours}
\end{figure}

In the limit $T\to 0$, $\nf(x) = \Theta(-x)$, so that $N(\bp)$ exhibits discontinuities whenever
an argument of one of the Fermi functions vanishes.  We proceed by choosing the FFLO wavevector
to be along the $\xh$ axis, $\bq = q\xh$ (valid since $S$ is independent of this choice), and
define  $p_+(\theta)$ and $p_-(\theta)$
to be the solutions to $\xi_{\bp-\frac{1}{2}\bq \uparrow}=0$ and  $\xi_{\bp+\frac{1}{2}\bq \downarrow}=0$, 
with $\theta$ the angle between $\bp$ and $\bq$.  We find:
\be
\label{Eq:ppm}
p_{\pm}(\theta)  = \frac{1}{2}\big(\sqrt{8(\mut\pm h)+q^2\cos^2\theta}\pm q\cos\theta\big) .
\ee
As we will show, the behavior of Eq.~(\ref{Eq:simpler}) depends crucially
on whether the circles $p_+(\theta)$ and $p_-(\theta)$ 
(that we can regard as the Fermi surfaces, \lq\lq boosted\rq\rq\ by $\pm \frac{1}{2}\bq$) 
intersect.  The two panels in Fig.~\ref{fig:contours}
show these circles for parameters such that $q<q_c$ (left panel) and $q>q_c$ (right panel) where 
\be
\label{eq:qcdef}
q_c \equiv\sqrt{2(\mu+h)}-\sqrt{2(\mu-h)},
\ee
 is the difference in Fermi wavevectors of the two species.  The yellow dashed line in these plots
shows where the denominator in  Eq.~(\ref{Eq:simpler}) vanishes.

In the next section, we show that the FFLO phase transition occurs for $q=q_c$ and, subsequently, we show that
the stable FFLO phase  always occurs for $q>q_c$.  To see this from a physical perspective, we note the yellow dashed line in the left and right panels
of Fig.~\ref{fig:contours}, which shows the curve, $p^2 = 2\mut$, where the denominator of Eq.~(\ref{Eq:simpler}) vanishes. 
This vanishing denominator represents a set of gapless excitations that are susceptibile to pairing (as in the usual
Fermi surface of a balanced fermion gas).
 However, for $q<q_c$, the curves $p_{\pm}$ never cross
the yellow dashed line, indicating that the numerator $N(\bp)$ always equals zero for $\bp$ on the yellow dashed line.
For $q>q_c$,  the curves $p_{\pm}$ cross each other at the same point that they cross the yellow dashed line, indicating the
presence of low energy fermionic excitations that are susceptible to pairing.
From this
qualitative picture, we thus expect pairing to occur for $q>q_c$, and to be strongest near these points of intersection.

\subsection{Jensen formula}

In the preceding subsection, we claimed that, at the FFLO phase transition, $q=q_c$.  To show
this, we must evaluate the integral in Eq.~(\ref{Eq:simpler}).  Although this integral can be evaluated via
contour integration, here we use the Jensen formula from 
complex analysis~\cite{Ahlfors}, which states that, for $f(z)$ analytic, 
\be
\frac{1}{2\pi} \int_0^{2\pi} d\theta \, \ln |f({\rm e}^{i\theta})|
 = \ln |f(0)| +\sum_{i=1}^n \ln \frac{1}{|a_i|},
\label{Eq:jensen}
\ee
with $|a_i|$ the zeroes of $f(z)$ inside the unit circle $|z|<1$.   To use the Jensen formula, we
first must evaluate the radial momentum $p$ integral.  This can be written as
\bea
\nonumber
&&S(h,q,0) = \frac{1}{4\pi^2}\int_0^{2\pi} d\theta 
\int_0^\infty dp\, p \Big[\frac{1}{p^2-2\mut} - \frac{1}{p^2+2\epsilon_b}
\\
&&
 \qquad -  \frac{\Theta(p_+(\theta)-p)}{p^2-2\mut} -\frac{\Theta(p_-(\theta)-p)}{p^2-2\mut} \Big].
\eea
We evaluate the $p$ integral, treating the divergence at $p= \sqrt{2\mut}$ within principal value, obtaining:
\be
S(h,q,0) = -\frac{1}{8\pi^2} \int_0^{2\pi} d\theta \, \ln \frac{|(p_+^2(\theta)-2\mut)(p_-^2(\theta)-2\mut)|}{2\mut \epsilon_b}.
\ee
To express this in the form of Eq.~(\ref{Eq:jensen}) (albeit with a different overall prefactor), we define $p_{\pm}(z)$ 
by replacing $\cos\theta \to \frac{1}{2}\big(z+\frac{1}{z}\big)$ in Eq.~(\ref{Eq:ppm}).  Then, we can write $S(h,q,0)$ as
\bea
S(h,q,0) &=&- \frac{1}{8\pi^2}
\int_0^{2\pi} d\theta \,\ln |f\big({\rm e}^{i\theta})|,
\label{Eq:oftheform}
\\
f(z) &\equiv & \frac{z^2( p_+^2(z)-2\mut)(p_-^2(z)-2\mut)}{2\mut \epsilon_b}.
\eea
Note the factor $z^2$ in  $f(z)$.  This factor is allowed, since under
the replacement $z\to {\rm e}^{i\theta}$, it will only give a factor $|z^2| =1$
in the argument of the logarithm.  Additionally, with this factor 
$f(z)$ does not have a pole at $z\to 0$.  To use the Jensen formula, we need
$|f(0)| = \frac{1}{2}\mut q^2$ and 
the zeros of $f(z)$, which are determined by the solutions to $ p_+^2(z)=2\mut$ and $p_-^2(z) = 2\mut$.
These are:
\be
z_\pm = \frac{-h \pm \sqrt{h^2- h_c^2}}{h_c} ,
\ee
where $h_c = q\sqrt{\mut/2}$.  For $h<h_c$ (equivalent to $q>q_c$), the zeros $z_\pm$ are on the unit  
circle in the complex plane (i.e. $|z_{\pm}|=1$), so that they give a vanishing contribution to the
sum in Eq.~(\ref{Eq:jensen}).  For $h>h_c$  (equivalent to $q<q_c$), only $z_+$ is inside the unit circle
(i.e., $|z_-|>1$ and $|z_+|<1$).  Since it is a double zero of $f(z)$, it contributes twice to 
the sum in Eq.~(\ref{Eq:jensen}).  Combining all terms gives:
\be
S(h,q,0) = \begin{cases}
\frac{1}{2\pi} \ln \frac{\sqrt{2\mut \epsilon_b}}{h+ \sqrt{h^2-\frac{1}{2}q^2\mut} }
 &\text{for} \,\,  q<q_c, \,\, \cr
\frac{1}{4\pi} \ln \frac{4\epsilon_b}{q^2} 
 & \text{for} \,\,  q>q_c.\,\, 
\end{cases}
\label{eq:essfflo}
\ee
Remarkably, for $q>q_c$ (the regime where the curves $p_{\pm}$ cross in Fig.~\ref{fig:contours}), 
 $S(h,q,0)$ is {\em independent} of $\mu$ and $h$.  Below, we show
that this simple result also holds for the full integral $S(h,q,\Delta)$ for sufficiently small $\Delta$.
Before doing this, we first use Eq.~(\ref{eq:essfflo}) to find the location of the FFLO phase boundary. 

\begin{figure}[ht!]\vspace{-.25cm}
     \begin{center}
        \subfigure{
            \includegraphics[width=85mm]{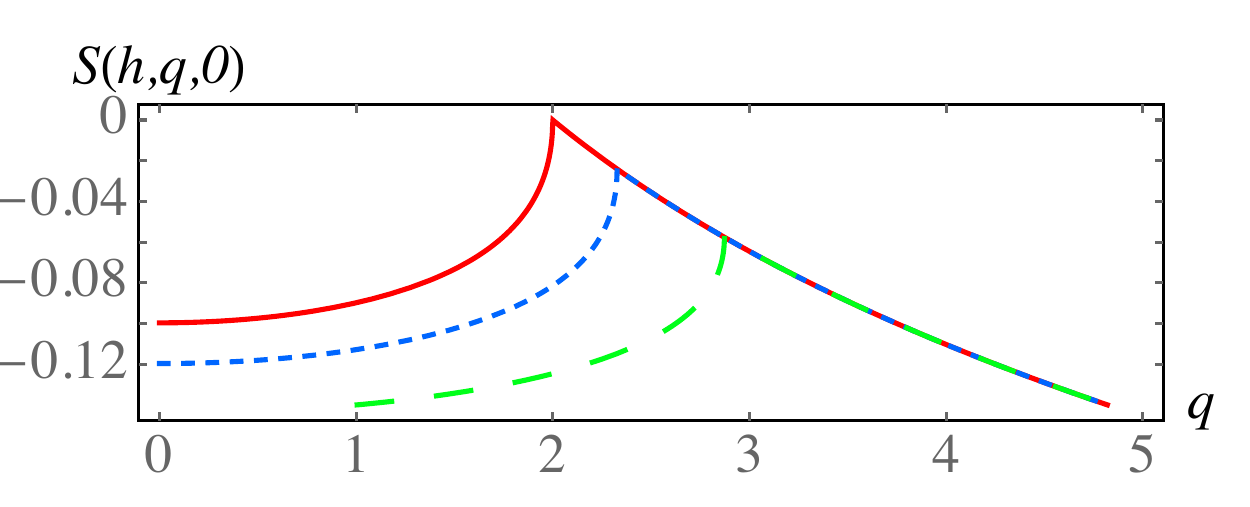}}
        \\ \vspace{-.55cm}
        \subfigure{
            \includegraphics[width=85mm]{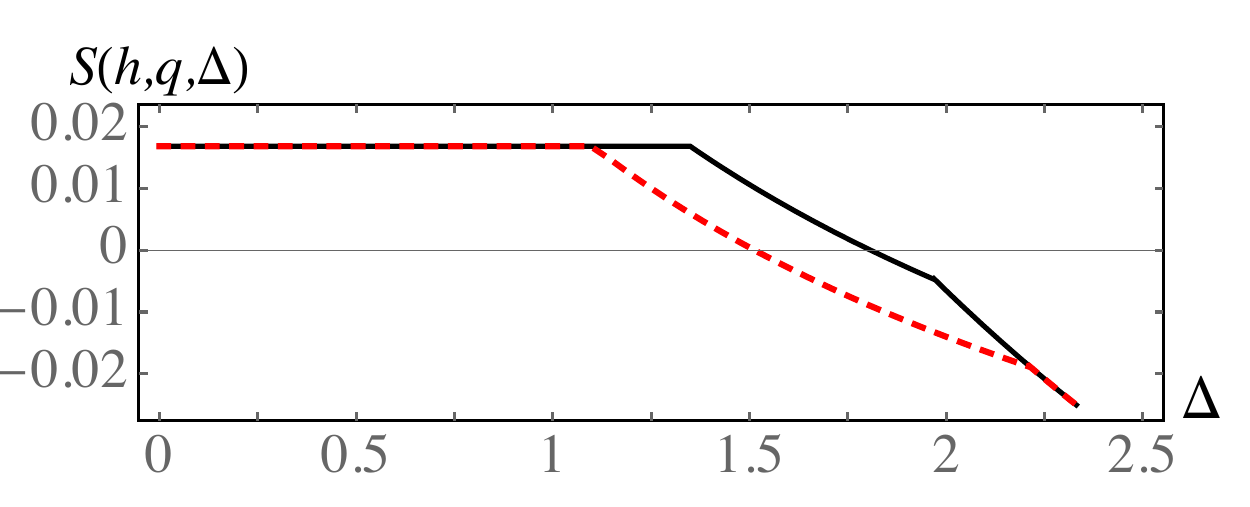}}
    \end{center}\vspace{-.5cm}
    \caption{
        (Color Online) The top panel shows $S(h,q,\Delta)$ at $\Delta = 0$ for $\mu = 4$, $\epsilon_b = 1$, and 
$h =3.5$ (green long-dashed), $h=3$  (blue short-dashed), and $h=\sqrt{7}\simeq 2.65$ (red solid), with the latter showing a finite
$q$ pairing instability where $S(h,q,0)=0$.  The bottom panel shows $S(h,q,\Delta)$ as a function of $\Delta$
for $\mut = 1.3$, $q = 1.8$ and $h=0.5$ (dashed red) and $h=0.28$ (solid black), with the stable FFLO phase
occuring at $S(h,q,\Delta)=0$.  The kinks in $S(h,q,\Delta)$
occur at $\Delta_{c1}$ and $\Delta_{c2}$, and a stable gap equation solution (minimum of the energy) satisfies $S(h,q,\Delta) = 0$.
A central question, then, is how to find a sensible approximation to $S(h,q,\Delta)$ given the unusual $\Delta$ dependence of this 
quantity. 
}
   \label{fig:ess}
\end{figure}

\subsection{Calculation of FFLO phase boundary}
In Fig.~\ref{fig:ess} (top panel), we plot $S(h,q,0)$ for $\mu = 4$ and for 
three values of $h$, using units such that $\epsilon_b\equiv 1$.  The kink in each curve occurs
at $q= q_c$, and, consistent with Eq.~(\ref{eq:essfflo}), the curves overlap for $q>q_c$.  To interpret
these physically, we note that the normal phase is stable against a continuous transition if $S(h,q,0)<0$.
Therefore, since the maximum of this curve is the kink location (at $q_c$) we see that the FFLO phase occurs
at $q = q_c(\mu,\hfflo)$ and $S(\hfflo,q,0)=0$, the simultaneous solution of which is 
\bea
\hfflo &=& \sqrt{2\mu \epsilon_b-\epsilon_b^2},
\\
q_c &=& 2\sqrt{\epsilon_b},
\label{Eq:qceetransition}
\eea
with the latter formula denoting the universal FFLO wavevector at the transition discussed above.  Thus, we find that, at
the FFLO transition, the momentum-shifted spin-$\uparrow$ and spin-$\downarrow$ Fermi surfaces touch at exactly
one point.  In the FFLO phase occuring for $h<\hfflo$, we will always have $q>q_c$ [as in the right panel of 
Eq.~(\ref{fig:contours})].

In Fig.~\ref{fig:one} (top panel), we plot the FFLO phase boundary $\hfflo$ as a red curve.  For sufficiently
small $\mu$, it crosses $h_c(\mu)$ at $\mu =  5\epsilon_b/4$, indicated as a blue point in this figure.  
To the left of this point, the FFLO phase transition from the normal phase, occuring with decreasing $h$,
is preceded by the first-order N-SF phase transition (so the regime of FFLO vanishes).  To the right of this point, the
phase boundary $h_c$, derived above as the N-SF phase boundary,  becomes a SF-FFLO first order transition.  Since
the FFLO is only slightly lower in energy than the N phase, the true $h_c$ will be only slightly lower than that 
depiced in Fig.~\ref{fig:one} (top panel).

The red curve labeled $\hfflo$ in Fig.~\ref{fig:one} (bottom panel) is the location of the continuous FFLO phase 
transition at fixed total particle number.  To obtain this, we simply need to use the chemical potential 
in the imbalanced normal state $\mu = \ef$.  This gives $\hfflo = \epsilon_b \sqrt{2\ef/\epsilon_b -1}$.  
Finally, we derive the critical polarization for the FFLO phase in the fixed magnetization and fixed
total density ensemble.  Using that the imbalanced normal state magnetization is $M = n_\uparrow + n_\downarrow = h/\pi$,
we obtain $\pfflo = \frac{\epsilon_b}{\ef}\sqrt{2\frac{\ef}{\epsilon_b}-1}$, plotted in Fig.~\ref{fig:PCphase}.

\begin{figure}[ht!]
     \begin{center}
        \subfigure{
            \includegraphics[width=41mm]{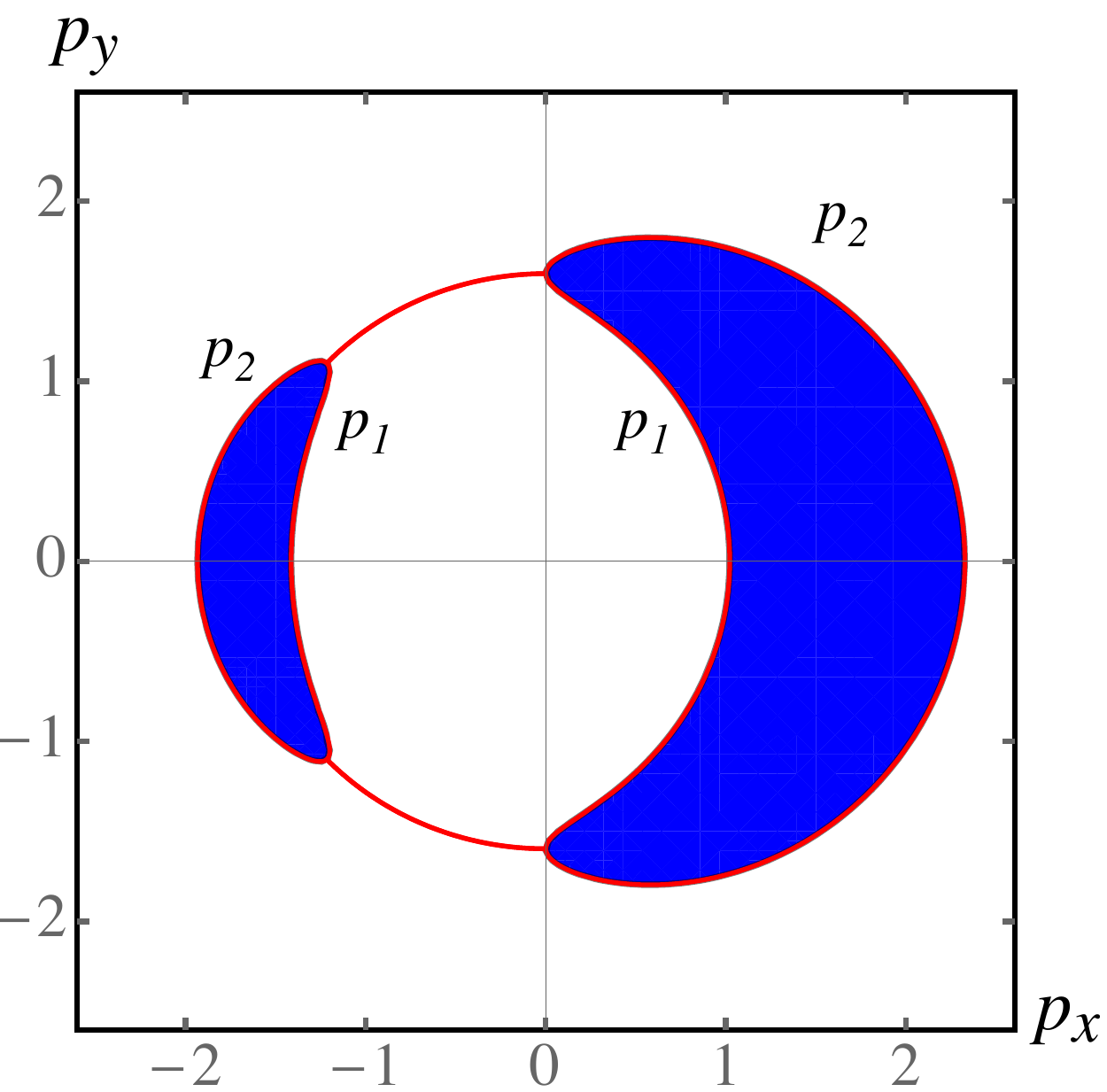}}
        \subfigure{
            \includegraphics[width=41mm]{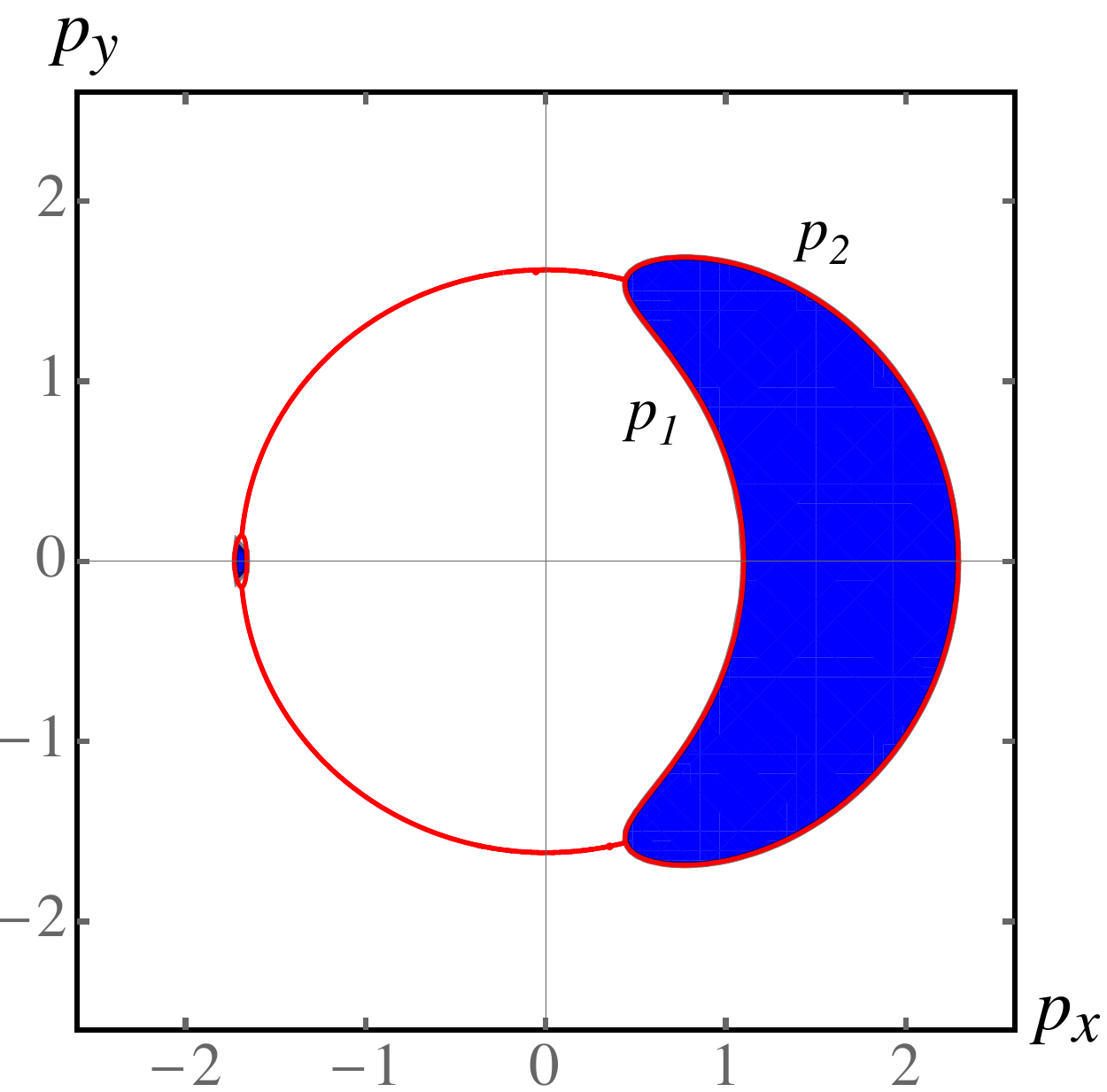}}
    \end{center}\vspace{-.5cm}
    \caption{
        (Color Online)  The left panel shows where $E_{\bp+}<0$ (left blue-shaded region)
and $E_{\bp-}<0$ (right blue shaded region), with parameters chosen so that $\Delta<\Delta_{c1}$.  The right panel
is the same, but with larger $\Delta$ such that $\Delta\alt \Delta_{c1}$.  We chose $\mu=1.4$, $q=1$, and $h=0.3$ for 
each (in units such that $\epsilon_b=1$).  For these parameters, $\Delta_{c1}\simeq 0.523$.  The left panel is for $\Delta=0.3$ 
and the right panel is for $\Delta=0.5$.
     }\vspace{-.25cm}
   \label{fig:contoursD}
\end{figure}

To conclude this section, we have shown that,  for $\Delta =0$, the integral controlling the gap equation $S(h,q,0)$ has
a nonanalytic dependence on $q$ that followed from the Jensen formula of complex analysis (but which is difficult to obtain
otherwise).  In the next section, we show that $S(h,q,\Delta)$ also possesses nonanalyticities at nonzero $\Delta$, 
complicating the problem of determining the stable pairing amplitude in the FFLO phase. 

\section{Analysis of $S(h,q,\Delta)$}

In the preceding sections, we analyzed the phase diagram of 2D imbalanced Fermi gases and found the phase boundary
for the onset of the FFLO phase.  Our analysis relied on calculating the gap equation integral, $S(h,q,\Delta)$, in
the limit $\Delta\to 0$, finding that $S(h,q,0)$ is independent of $\mu$ and $h$ for sufficiently large $q$. 

To understand the strength of pairing in the FFLO state below $\hfflo$, in this section
we analyze the $\Delta$ dependence of $S(h,q,\Delta)$.  We will show that $S(h,q,\Delta)$ satisfies,
for $q>q_c$ and $T\to 0$, 
\be
S(h,q,\Delta) = \begin{cases}   \frac{1}{4\pi}
\ln \frac{4\epsilon_b}{q^2}
  & \text{for} \,\, \Delta<\Delta_{c1}, \cr
\frac{1}{4\pi}\ln \frac{\epsilon_b}{\sqrt{\Delta^2 +\mut^2}-\mut}
  & \text{for} \,\, \Delta>\Delta_{c2},
\end{cases}
\label{Eq:essphenomSM}
\ee
with $\Delta_{c1}$ and $\Delta_{c2}$ defined below.  For clarity, we emphasize here the $q_c$ is defined by
Eq.~(\ref{eq:qcdef}).  In contrast, Eq.~(\ref{Eq:qceetransition}) denotes the value of $q_c$ at the FFLO phase
transition.  

In Fig.~\ref{fig:ess} (bottom panel) we show $S(h,q,\Delta)$ as a function of $\Delta$ for two different typical parameter
choices (the black curve and the red curve), obtained by numerically integrating the integral governing $S$.
 Remarkably, consistent with the first line of Eq.~(\ref{Eq:essphenomSM}), $S$ is indeed independent of $\Delta$ for small
$\Delta$, and equal to the second line of Eq.~(\ref{eq:essfflo}).  The two slope discontinuities (or kinks) in these
curves are at $\Delta_{c1}$ and $\Delta_{c2}$, and we find  that 
the stable FFLO phase always occurs for $\Delta_{c1}<\Delta<\Delta_{c2}$.
Although we have not found an analytic form for $S(h,q,\Delta)$ in this regime, our method (based on the Jensen formula) 
for demonstrating
the behavior in Eq.~(\ref{Eq:essphenomSM}) leads to a useful analytical approximation to $S(h,q,\Delta)$ valid for $\Delta\agt \Delta_{c1}$
(that we present in Sec.~\ref{sec:as}).

We start by directly evaluating some of the integrals in Eq.~(\ref{Eq:fulldelta}).  We find:
\bea
\label{Eq:fulldeltap}
&&\hspace{-0.75cm}S(h,q,\Delta)
=  S_0 + S_1,
\\
&&\hspace{-0.75cm}S_0 \equiv   \frac{1}{4\pi} \ln \frac{\epsilon_b}{\sqrt{\Delta^2+\mut^2}-\mut} ,
\label{Eq:essnought}
\\
&&\hspace{-0.75cm}S_1 \equiv    
- \frac{1}{4\pi^2}\int_0^{2\pi}d\theta\int_0^\infty dp\,p 
\frac{\nf(E_{\bp+}) + \nf(E_{\bp-})}{2E_p},
\label{eq:essone}
\eea
where $S_1$ includes the terms with Fermi functions.
We now analyze $S_1$ in the limit of $T\to 0$, where  $\nf(x) \to \Theta(-x)$, so that the integrand is only 
nonzero for momenta such that $E_{\bp\pm}<0$.  To proceed, we need to find the solutions to 
$E_{\bp\pm}= 0$, which are 
\bea
 0 &=& E_p \pm \big( h + \frac{\bp \cdot \bq}{2m}\big),  
\eea
which, after moving $E_p$ to one side, squaring both sides, simplifying and setting $m=1$, takes
the form of a quartic equation~\cite{Abramowitz}:
\be
\label{Eq:fullquartic}
p^4 - (4\mut+q^2\cos^2\theta)p^2-4hq\cos\theta p + 4(\Delta^2-h^2+\mut^2) = 0 .
\ee
Here, we took $\bq$ along the positive $x$ axis with $\theta$ the angle between $\bp$ and $\bq$. The solution 
to the quartic equation is rather complex, and we relegate some of the details to 
Appendix~\ref{sec:appquar}.  Briefly, although there are generally four solutions to the quartic equation, in the
present case there are at most two real and positive solutions, which are given by:
\bse
\label{Eq:pees}
\bea
 p_2(\theta) &=& \Sh + \frac{1}{2}\sqrt{-4\Sh^2 -2p-\frac{\qh}{\Sh}},
\\
p_1(\theta) &=& \Sh - \frac{1}{2}\sqrt{-4\Sh^2 -2p-\frac{\qh}{\Sh}},
\eea
\ese
defining two momenta that, when they are real, define the boundaries of the regions where
$E_{\bp\pm}<0$, as illustrated in Fig.~\ref{fig:contoursD}.  Here, $p$, $\qh$ and $\Sh$  are defined in Appendix~\ref{sec:appquar},
with the hats on $\qh$ and $\Sh$ intended to distinguish them from the quantities $q$ and $S$.

We now evaluate the radial $p$ integral of Eq.~(\ref{eq:essone}) at $T=0$ for a particular angle $\theta$.  
There are two possible cases to consider.  In the first case, for angles $\theta$
such that the radial $p$ integration intersects one of the regions where $E_{\bp\pm}<0$, the Fermi
functions effectively  restrict the integration range to $p_1<p<p_2$.  
In the second case,  for angles $\theta$ such
that the radial $p$ integration does not intersect a region where $E_{\bp\pm}<0$, the integrand
vanishes.  In both cases, the result of the radial $p$ integral gives:
\be
S_1  = \frac{1}{8\pi^2} \int_0^{2\pi}d\theta \ln \Big|\frac{p_1^2 -2\mut + 2E_{p_1}}
{p_2^2 -2\mut + 2E_{p_2}}\Big|.
\label{Eq:essonefinal}
\ee
 This follows because, when $p_1$ and $p_2$ are real,
the integrand comes from the radial $p$ integral corresponding to the first case.  In the second case, $p_1$ and
$p_2$ are complex quartic equation solutions (that are complex conjugates of each other), 
and the absolute value bars inside the logarithm ensure that the argument of the logarithm 
is unity, giving zero for the integrand.   

In the next subsection we demonstrate the second line of Eq.~(\ref{Eq:essphenomSM}).

\subsection{Case of $\Delta>\Delta_{c2}$}
The form of Eq.~(\ref{Eq:essonefinal}) suggests that we may apply the Jensen formula Eq.~(\ref{Eq:jensen}) if
the argument of the logarithm is analytic.  We will show that this function is analytic for $\Delta<\Delta_{c1}$
(allowing us to use the Jensen formula), 
but that a branch cut appears for $\Delta>\Delta_{c1}$.
Here, $\Delta_{c1}$ and $\Delta_{c2}$ (obtained below) are defined by the values of $\Delta$ at
which the regions of $E_{\bp\pm}<0$ shrink to zero.  In particular, for $\Delta<\Delta_{c1}$, the regions
where  $E_{\bp+}<0$ and $E_{\bp-}<0$ are given by the kidney shaped blue regions plotted in Fig.~\ref{fig:contoursD}.
Note that the leftmost region (where $E_{\bp+}<0$) is smaller than the rightmost region (where $E_{\bp-}<0$).

At $\Delta_{c1}$, the region 
where $E_{\bp+}<0$ disappears, as depicted in the right panel of Fig.~\ref{fig:contoursD} (which is for
parameters such that $\Delta$ is slightly below $\Delta_{c1}$).
For $\Delta_{c1}<\Delta<\Delta_{c2}$, only the rightmost blue shaded region, where $E_{\bp-}<0$ 
exists.  It shrinks with increasing $\Delta$ until, at $\Delta_{c2}$, the region where $E_{\bp-}<0$
disappears.

Thus, for the case of $\Delta>\Delta_{c2}$, both Fermi functions in the numerator of Eq.~(\ref{eq:essone}) vanish
and $S_1=0$.  This immediately yields the second line of Eq.~(\ref{Eq:essphenomSM}).
  Thus, our remaining tasks are
to derive the behavior in the regime $\Delta<\Delta_{c1}$ [the first line of Eq.~(\ref{Eq:essphenomSM})] and to
obtain an expression for $\Delta_{c1}$ and $\Delta_{c2}$.

\subsection{Case of $\Delta<\Delta_{c1}$}
Turning to the regime of $\Delta<\Delta_{c1}$, 
we first rewrite Eq.~(\ref{Eq:essonefinal}) using Eq.~(\ref{eq:eepee}):
\be
S_1 = \frac{1}{8\pi^2}\int_0^{2\pi} d\theta\, 
\ln \Big|\frac{(p_1^2-2\mut + 2E_{p_1})(p_2^2-2\mut - 2E_{p_2})}{4\Delta^2}\Big|,
\label{Eq:logarg}
\ee
which, defining $z = {\rm e}^{i\theta}$, can be written as 
\bea
&&\hspace{-0.75cm}S_1 = \frac{1}{8\pi^2}\int_0^{2\pi} d\theta\, \ln |f({\rm e}^{i\theta})|, 
\\
&&\hspace{-0.75cm}f(z) \!=\!\frac{(p_1^2(z)\!-\!2\mut\! +\! 2E_{p_1(z)})(2E_{p_2(z)} \!-\! p_2^2(z)\!+\!2\mut )}{4\Delta^2z^2}.
\label{Eq:effgen}
\eea
Note that we included a factor of $z^2$ in the denominator, which is mathematically correct since the integral is 
on the complex unit circle $|z^2|=1$.  The reason for including this factor is it simplifies the usage of the Jensen formula, which relies
on understanding the behavior of $f(z)$ inside the unit circle.   In this expression, 
$p_1(z)$ and $p_2(z)$ are defined by making the replacement
\be
\label{eq:replacement}
\cos\theta \to \frac{1}{2}\big(z+\frac{1}{z}\big),
\ee
in the quantities appearing in Eq.~(\ref{Eq:pees}).

The function $f(z)$ has no zeros inside the unit circle, so to use the Jensen formula we only need $f(0)$.   This in turn requires 
$p_1(z)$ and $p_2(z)$ for $z\to 0$, which are most easily obtained by returning to the original quartic
equation in the limit $z\to 0$.  Making the replacement Eq.~(\ref{eq:replacement}) in
Eq.~(\ref{Eq:fullquartic}), in the limit $z\to 0$ we have solutions of the form $p\propto az$ or $p\propto b/z$.
Plugging in these solutions, Taylor expanding to leading order in small $z$, and solving for $p$ yields for $a$ and $b$:
\bse
\label{eq:parameters4J}
\bea
a&=&\frac{4}{q} \big(-h\pm \sqrt{\Delta^2 + \mut^2}) ,
\\
b&=& \pm \frac{q}{2},
\eea
\ese
describing four solutions to our quartic equation for $z\to 0$. We need to know which of these solutions corresponds
to our solutions, $p_1(z)$ and $p_2(z)$, in the limit $z\to 0$.  To do this, for simplicity, we can approach $z=0$ along the 
positive real axis, so that  all four solutions are real.  From the general theory of the quartic equation, 
it is known that the properties of quartic equation solutions can be characterized by the
 discriminant $\Delta_d$ [defined in Eq.~(\ref{eq:discriminant})] and the quantity $P$ defined in 
Eq.~(\ref{Eq:bigpee}). 
  The latter is given by (upon using Eq.~(\ref{eq:replacement})),
\bea
P= -32\mut - 2q^2\big(\frac{1}{z}+z\big)^2,
\eea
and is negative for $z$ real and positive.  Similarly, the discriminant, although rather complicated, has
the limiting form
\bea
\Delta_d \simeq \frac{q^8}{4z^8} \big(\Delta^2 + \mut^8\big),
\eea
for $z\to 0$
and is positive for $z$ real and positive.  When  $\Delta_d>0$ and $P<0$, as is the case here, the
 quartic equation solutions are either
all complex or all real.  Clearly, the present case is the latter, with the explicit form of the four
quartic solutions given in Eq.~(\ref{Eq:generalquartic}), where the parameter $b=0$ and $\Sh>0$.
Since the solutions $x_{1,2}$ are negative, 
the analytical continuation of $p_1$ and $p_2$ are the two + cases in Eqs.~(\ref{eq:parameters4J}).  Furthermore,
since  $p_2>p_1$, we have:
\bea
&&p_2 \sim \frac{q}{2z},
\\
&&p_1 \sim \frac{4}{q} \big(-h+\sqrt{\Delta^2 + \mut^2}) z.
\eea
The preceding arguments showed how the quartic equation solutions 
 behave under the replacement  Eq.~(\ref{eq:replacement}).  Now all that remains is to plug
these into Eq.~(\ref{Eq:effgen}) and take the limit $z\to 0$.  We find:
\be
f(0) =  \frac{4(\sqrt{\Delta^2 + \mut^2}-\mut)}{q^2},
\ee
which, using Eq.~(\ref{Eq:jensen}), leads to:
\be
S_1  = \frac{1}{4\pi} \ln \frac{4(\sqrt{\Delta^2 + \mut^2} - \mut)}{q^2},
\ee
and $S(h,q,\Delta) = \frac{1}{4\pi} \ln \frac{4\epsilon_b}{q^2}$, demonstrating 
the first line of Eq.~(\ref{Eq:essphenomSM}).


This, as described above, for small $\Delta$ the gap equation integral is {\em independent\/} of $\Delta$
(clearly invalidating using a Taylor expansion as a strategy to approximately evaluate $S(h,q,\Delta)$).
Our remaining tasks are to determine the critical pairing amplitudes $\Delta_{c1}$ and $\Delta_{c2}$ and
to show that a branch cut in Eq.~(\ref{Eq:effgen}) appears for $\Delta>\Delta_{c1}$.

\subsection{Critical pairing amplitudes $\Delta_{c1}$ and $\Delta_{c2}$}
\label{Eq:cpa}

In this section, we will relate the critical pairing amplitudes $\Delta_{c1}$ and $\Delta_{c2}$ 
to the quartic equation discriminant $\Delta_d(\theta)$. 
The discriminant (discussed in Appendix~\ref{sec:appquar}) helps classify properties of quartic equation solutions.  Although it is
defined for any angle $\theta$ (since Eq.~(\ref{Eq:fullquartic}) is defined for any $\theta$), in this section we will be mainly interested
in the discriminant for $\theta = 0$. 
  As shown below, and illustrated in Fig.~\ref{fig:disc}, $\Delta_d(0)$ vanishes at $\Delta_{c1}$ and $\Delta_{c2}$.  We find
this to be the simplest way to determine these quantities.

 To define  $\Delta_{c1}$ and $\Delta_{c2}$, consider parameter
ranges in which both of $E_{\bp+}$ and $E_{\bp-}$ are negative for some $\bp$.  In Fig.~\ref{fig:contoursD} (left panel)
we depict such a case, with $E_{\bp+}<0$ in the left blue region and $E_{\bp-}<0$ in the right blue region.  With
increasing $\Delta$ (holding other parameters fixed), the leftmost blue region vanishes at $\Delta_{c1}$ and
the rightmost blue region vanishes at $\Delta_{c2}$.  

\begin{figure}[ht!]\vspace{-.25cm}
     \begin{center}
%
            \includegraphics[width=85mm]{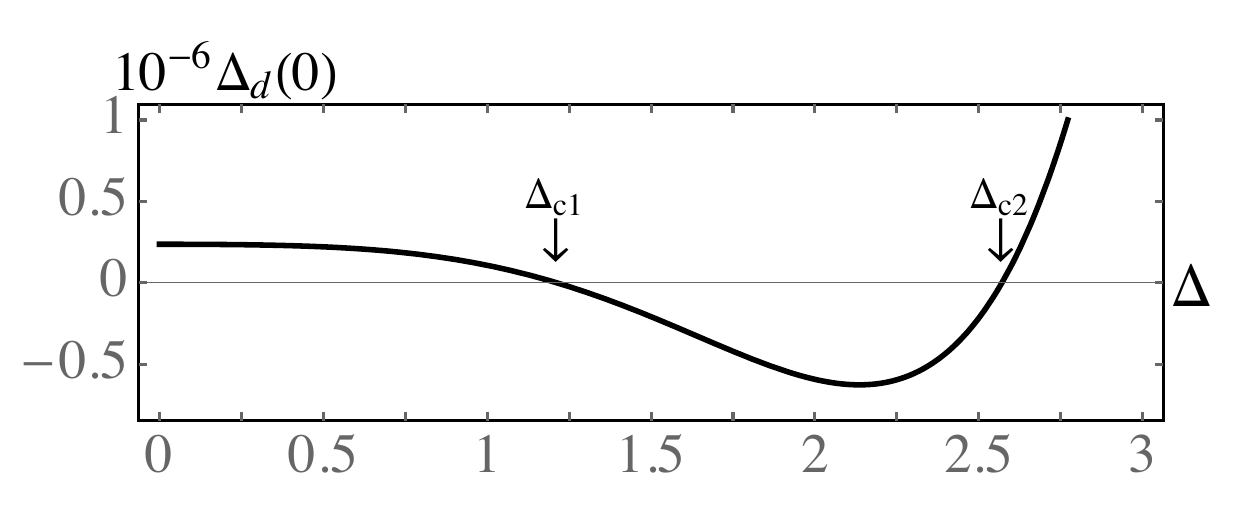}
    \end{center}\vspace{-.5cm}
    \caption{Plot of the quartic equation discriminant, $\Delta_d(\theta)$, at angle $\theta =0$
(and multiplied by $10^{-6}$) for $\mut=1.3$, $h = 0.6$ and $q=2$ and as a function of
$\Delta$.  As shown in
Sec.~\ref{Eq:cpa}, the qualitative behavior shown here, in which $\Delta_d(0)$ vanishes
at $\Delta_{c1}$ and $\Delta_{c2}$, always occurs in the FFLO regime.  Therefore, we can determine
 $\Delta_{c1}$ and $\Delta_{c2}$ by solving $\Delta_d(0) = 0$ for the pairing amplitude $\Delta$. 
       }
   \label{fig:disc}
\end{figure}

We now relate $\Delta_{c1}$ and $\Delta_{c2}$ to our quartic equation solutions. 
Since the edges of the regions where $E_{\bp+}<0$ and $E_{\bp-}<0$ are determined by $p_1(\theta) = p_2(\theta)$,
to find $\Delta_{c1}$ we need to solve $p_1(\theta) = p_2(\theta)$ at $\theta = \pi$.  Similarly, $\Delta_{c2}$ is 
obtained by solving  $p_1(\theta) = p_2(\theta)$ at $\theta = 0$.

Examining Eq.~(\ref{Eq:pees}), we see that the condition $p_1(\theta) = p_2(\theta)$ implies the vanishing of
the argument of the square root,  or $R_-(\theta) =0$ where 
\bea
\label{eqrmp}
R_\pm(\theta) &=& -4\Sh^2 -2p\pm\frac{\qh}{\Sh}
\\
 &=& -4\Sh^2 + 2q^2 \cos^2\theta +8\mut  \mp \frac{4\cos\theta hq}{\Sh},
\label{eqrmp2}
\eea
where we emphasize that $p$, $\qh$ and $\Sh$ all depend on $\theta$.  
Here, $R_-$ appears in the quartic equation solutions defining the boundaries where $E_{\bp\pm}$ cross zero,
and $R_+$ appears in the other two quartic equation solutions.  

The preceding discussion shows that the region where $E_{\bp+}<0$ vanishes when $R_-(\pi) = 0$, defining $\Delta_{c1}$.
With further increasing $\Delta$, the region where $E_{\bp-}<0$ vanishes when $R_-(0) = 0$, defining $\Delta_{c2}$.  However, since the
function $\Sh$ depends only on $\cos^2\theta$ (so that it is the same at $\theta=0$ and
at $\theta  = \pi$),  Eq.~(\ref{eqrmp2}) implies that $\Delta_{c2}$ is equivalently defined by $R_+(\pi) = 0$.  Thus, $\Delta_{c1}$ 
is defined by $R_-(\pi) = R_+(0) = 0$ and $\Delta_{c2}$ is defined by $R_-(0) = R_+(\pi) = 0$.

We can furthermore
connect $R_{\pm}(\theta)$ to the discriminant  $\Delta_d(\theta)$, using Eq.~(\ref{Eq:discsols}) of
the Appendix that relates $\Delta_d(\theta)$ to the four quartic solutions. 
This equation is
\be
\label{Eq:discsolsmain}
\Delta_d = \sum_{i<j} (x_i - x_j)^2,
\ee
with the $x_i$ ($i=1,\cdots 4$) being the four quartic solutions.  By plugging in our quartic solutions, 
and simplifying, we find
\be
\label{deltadrpm}
\Delta_d(\theta) = R_+(\theta)R_-(\theta)\frac{1}{16 \Sh^4} \big( 128 \Sh^6 +32\Sh^4 p +\qh^2)^2 ,
\ee
showing that $\Delta_d(\theta)$ also must vanish, for $\theta = 0$ and $\theta = \pi$, at $\Delta_{c1}$ and $\Delta_{c2}$.  Indeed, 
by examining Eq.~(\ref{eq:discriminant}) of Appendix~\ref{sec:appquar},
 which relates $\Delta_d(\theta)$ to the quartic equation coefficients via
the functions $\Delta_0$ and $\Delta_1$, we see that $\Delta_d(\theta)$ is also only dependent on $\cos^2\theta$, so that it has
the same value at $\theta = 0$ and $\theta = \pi$. Thus, we need only consider $\Delta_d(0)$.

The explicit expression for the discriminant at $\theta = 0$ is:
\bea
&&\Delta_d(0) = 64 \Big(
h^2q^2(q^2+4\mut)^3-108h^4q^4
+\Deltat^2\big[ (q^2+4\mut)^4
\nonumber \\
&&
-144h^2q^2(q^2+4\mut)\big]
 + 32 \Deltat^4 (q^2+4\mut)^2 +256 \Deltat^6
\Big),
\eea
where we defined $\Deltat^2\equiv \Delta^2 - h^2+\mut^2$.  We obtain the critical pairing amplitudes
by solving $\Delta_d(0) = 0$, which is a cubic equation for $\Delta^2$.  Thus, we can find $\Delta_{c1}$
and $\Delta_{c2}$ by solving this cubic equation and taking the square root.  Since the cubic equation solution
is well known (and complicated), we will not reproduce the result but, instead, now  show that $\Delta_{c1}$
and $\Delta_{c2}$ are the {\em only} real solutions to $\Delta_d(0) = 0$. 

To do this, we note that, since the discriminant of the 
cubic equation for $\Delta^2$ is positive, it has three real solutions, two of which are $\Delta_{c1}^2$ and $\Delta_{c2}^2$.
However, since $\Delta_d(0)>0$ for $\Delta\to 0$ (for small $h$ in the FFLO regime), and $\Delta_d(0)\to +\infty$
for large $\Delta$, it is clear that there cannot be three real and positive solutions for $\Delta_d(0)=0$.  
The third solution to  $\Delta_d(0)$ must be for $\Delta^2<0$, and the two real and positive solutions to
$\Delta_d(0) =0$ determine $\Delta_{c1}^2$ and $\Delta_{c2}^2$.

Therefore we conclude that $\Delta_d(0)$ as a function of $\Delta$ generically looks like Fig.~\ref{fig:disc},
with $\Delta_d(0)>0$ for $\Delta<\Delta_{c1}$ and  $\Delta>\Delta_{c2}$ and $\Delta_d(0) <0$ for $\Delta<\Delta_{c1}$.
Although the full expressions for $\Delta_{c1}$ and $\Delta_{c2}$ are too unwieldy to show, they are easily obtained from
Eq.~(\ref{deltadrpm}) and the known cubic equation solutions.

\subsection{Branch cut for $\Delta>\Delta_{c1}$}

Our final task in this section is to show that a branch cut appears along the negative real axis for $\Delta>\Delta_{c1}$.  To do this, 
we consider our quartic equation solutions, $p_1(z)$ and $p_2(z)$, in the complex plane, that are obtained by replacing
$\cos\theta \to \frac{1}{2}\big(z+\frac{1}{z}\big)$ in Eq.~(\ref{Eq:pees}).  We also consider the generalized discriminant, $\Delta_d(z)$, 
given by:
\bea
\label{deltadrpma}
&&\hspace{-0.75cm}\Delta_d(z) = R_+(z)R_-(z)\frac{1}{16 \Sh^4} \big( 128 \Sh^6 +32\Sh^4 p +\qh^2)^2 ,
\\
&&\hspace{-0.75cm}R_\pm(z) = -4\Sh^2 + \frac{1}{2}q^2 \big(z+\frac{1}{z}\big)^2
 +8\mut  \mp \frac{2hq}{\Sh}\big(z+\frac{1}{z}\big),
\eea
by analogy with Eq.~(\ref{deltadrpm}).
We emphasize that $\Sh$, $p$, and $\qh$ all depend on $z$, although we have not displayed their dependence.  Since $p_1(z)$ and $p_2(z)$
depend on $R_-$ via $p_{1,2}(z) = \Sh \pm \frac{1}{2} \sqrt{R_-(z)}$, a vanishing of $R_-(z)$ for $z$ on the real axis implies
a branch cut in $p_{1,2}(z)$ (and therefore a branch cut in Eq.~(\ref{Eq:oftheform})), invalidating the use of the Jensen formula.   In this section
we show that this occurs for $\Delta>\Delta_{c1}$.

We start in the regime $\Delta<\Delta_{c1}$ and examine the behavior of $\Delta_d(z)$ on the real axis.  
Since the points $z=-1$  and $z=1$ correspond to $\theta = \pi$ and $\theta = 0$ on the unit circle, 
our previous discussion shows that $\Delta_d(z)>0$ at $z=\pm 1$ when $\Delta <\Delta_{c1}$.  
Furthermore,  since the $\theta$ dependence of $\Delta_d(\theta)$ enters only via $\cos^2\theta$, 
we know that the $z$ dependence of $\Delta_d(z)$ enters only via $\big(z+\frac{1}{z})^2$.  This 
implies that $\Delta_d(z)$ only has extrema at $z = \pm 1$.  Therefore, for $\Delta<\Delta_{c1}$,
$\Delta_d(z)>0$ for all $z$ on the real axis.  Along with the fact that  $P(z)$  in Eq.~(\ref{Eq:bigpee}) is negative, this
implies that the quartic equation has four real solutions, two of which are $p_1(z)$ and $p_2(z)$.  We conclude that no branch cuts
in $p_1(z)$ and $p_2(z)$ exist on the real axis in the range of interest $z=-1<z<1$.

Now consider $\Delta\geq \Delta_{c1}$.  Our previous analysis shows that $R_-(\theta)$ and the discriminant $\Delta_d(\theta)$ vanish for $\theta = \pi$ at $\Delta=\Delta_{c1}$,
and for $\Delta>\Delta_{c1}$  we found that $\Delta_d(\theta)$ is negative for $\theta = \pi$.  This implies that $\Delta_d(z)<0$ for $z$ in the vicinity of
$-1$ on the real axis, so that two of the four quartic solutions are complex.  The two complex solutions are $p_1(z)$ and $p_2(z)$, since the other
two solutions involve $R_+(z)$ that is positive.  Therefore, $R_-(z)<0$ for $z$ on the negative real axis, introducing 
a branch cut in Eq.~(\ref{Eq:oftheform}).  

To conclude this section, we see that, indeed, in the $T=0$ limit
 the gap equation integral Eq.~(\ref{Eq:fulldelta}) has the form displayed in Eq.~(\ref{eq:essfflo}), with the behavior
for $\Delta_{c1}<\Delta<\Delta_{c2}$ being unknown (analytically).  A numerical integration of Eq.~(\ref{Eq:fulldelta}) 
(plotted in Eq.~(\ref{fig:ess}))
confirms the form of 
$S(h,q,\Delta)$ for $\Delta<\Delta_{c1}$ and for $\Delta>\Delta_{c2}$, and also shows that $S(h,q,\Delta)$ has slope discontinuities 
at $\Delta_{c1}$ and $\Delta_{c2}$.

\section{Approximate Solution for $\Delta\agt\Delta_{c1}$}
\label{sec:as}

The non-analytic structure of $S(h,q,\Delta)$ precludes a Taylor expansion
in $\Delta$ for small $\Delta$ at zero temperature, since $S(h,q,\Delta)$ is exactly $\Delta$-independent for 
$\Delta<\Delta_{c1}$.  FFLO gap equation solutions, satisfying $S(h,q,\Delta)=0$, thus occur for $\Delta>\Delta_{c1}$.
We note that this does not imply a first order transition, since $\Delta_{c1}$ vanishes at $\hfflo$ (as we show below).  Indeed,
we find that  the transition is continuous, with $\Delta_{c1}$ increasing from zero as $h$ drops below $\hfflo$ while
maintaining $\Delta>\Delta_{c1}$.

To understand this behavior, and to find an approximate analytical expression for the degree of pairing in the FFLO phase,
in this section, we make an expansion in small $\Delta- \Delta_{c1}$ that holds 
close to the FFLO phase transition.  We start by finding an approximate form for $\Delta_{c1}$.  We first note
that  $E_{\bp +}$ only crosses zero for $q>q_c$, where we recall that $q_c$, defined in Eq.~(\ref{eq:qcdef}), 
is the difference in Fermi wavevectors of the two species.
Since $\Delta_{c1}$ is defined by the pairing amplitude at which the regime where $E_{\bp +}<0$ disappears, 
we see that $\Delta_{c1}$ must vanish for $q\to q_c^+$.  Note that, since $q = q_c$ at $\hfflo$, this is equivalent to
the assertion (made above) that $\Delta_{c1}\to 0$ for $h\to \hfflo$.  

We proceed by assuming $\Delta_{c1}$ can be Taylor expanded in 
small $(q-q_c)$ as $\Delta_{c1} = a(q-q_c) + b(q-q_c)^2+\cdots$, with $a$ and $b$ unknown coefficients, 
and plug this ansatz into $\Delta_d(0) = 0$ (the vanishing of the quartic equation discriminant).  Keeping only
the linear order term proportional to $(q-q_c)$, we find:
\be
\Delta_{c1} \simeq \frac{1}{\sqrt{2}} (\mu^2 - h^2)^{1/4}(q-q_c) .
\label{eq:deltac1nearkc}
\ee
Henceforth,  in the subsections below, we will use Eq.~(\ref{eq:deltac1nearkc}) for $\Delta_{c1}$. 

\subsection{Gap equation integral for $\Delta\agt\Delta_{c1}$}
 Our next task is to find an approximation 
for $S(h,q,\Delta)$ that is valid for $\Delta\agt\Delta_{c1}$.  
We do this in an indirect way that makes use of the Jensen formula result for the behavior of $S(h,q,\Delta)$, allowing us to 
infer the behavior of this quantity for $\Delta>\Delta_{c1}$ from the behavior for $\Delta<\Delta_{c1}$.

We start by writing Eq.~(\ref{eq:essone})
as $S_1 = S_+ + S_-$ with 
\bea
S_{\pm} &=& -\int \frac{d^2p}{(2\pi)^2} \frac{\nf(E_{\bp\pm})}{2E_p}.
\eea
For $\Delta \alt \Delta_{c1}$, the region where the integrand of $S_+$ is nonzero is about to shrink to zero, allowing us
to accurately approximate the integral in this limit.  The result is:
\be
\label{Eq:spluslast}
S_+ =   \frac{1}{4\pi h}\sqrt{\frac{q_c}{q-q_c}}\big(\Delta - \Delta_{c1}\big).
\ee
The calculations leading to Eq.~(\ref{Eq:spluslast}) are described in Appendix~\ref{sec:appsp}.

The result Eq.~(\ref{Eq:spluslast}) holds for $\Delta\alt \Delta_{c1}$.  However, our goal is to understand the 
region $\Delta\agt \Delta_{c1}$.   To determine this, we note that our previous analysis tells us that, for $\Delta<\Delta_{c1}$,
the full sum $S = S_0 + S_+ +S_-$ is 
\be
S = \frac{1}{4\pi}\ln \frac{4\epsilon_b}{q^2}.
\ee
Note that the fact that $S$ is $\Delta$-independent for $\Delta<\Delta_{c1}$ occurs despite the fact that $S_0$, $S_+$ and 
$S_-$ all depend on $\Delta$; this nontrivial behavior follows from the Jensen formula as described in the preceding section.

Using Eq.~(\ref{Eq:spluslast}), then, we obtain 
\be
\label{Eq:essminusares}
S_- = \frac{1}{4\pi} \ln\frac{4\epsilon_b}{q^2} -S_0 -  \frac{1}{4\pi h}\sqrt{\frac{q_c}{q-q_c}}\big(\Delta - \Delta_{c1}\big).
\ee
Note that, while $S_+$ vanishes for $\Delta>\Delta_{c1}$ (due to the Fermi function vanishing), $S_-$ does not.  In fact, $S_-$ should be a smooth function of $\Delta$ since its  
integrand, and the integration range, are smooth functions of $\Delta$.  This implies that Eq.~(\ref{Eq:essminusares}) should hold for
$\Delta\agt\Delta_{c1}$.  Combining this with the other terms determining $S$, we have 
\be
S(h,q,\Delta) = \frac{1}{4\pi}\ln \frac{4\epsilon_b}{q^2} - 
\frac{1}{4\pi h}\sqrt{\frac{q_c}{q-q_c}}\big(\Delta - \Delta_{c1}\big).
\label{closetodeltac}
\ee
Thus, although a direct approximate evaluation of $S_-$ would be difficult due to the shape of the integration region (e.g., the rightmost 
blue shaded regions of the left and right panels of Fig.~\ref{fig:contoursD}), the Jensen theorem allows us to connect this quantity
to $S_+$.  We are now able to obtain an approximate result for the stationary pairing amplitude, given by $S(h,q,\Delta) =0$:
\be
\Delta = \Delta_{c1}+ h \sqrt{\frac{\qb}{q_c}}\ln \frac{4\epsilon_b}{q^2} ,
\ee
where we defined $\qb = q-q_c$.  This equation must be solved along  with the stationarity condition for the FFLO wavevector $q$.
  To find the latter, 
we first determine the ground-state energy Eq.~(\ref{eq:eggeneral}) for $\Delta\agt \Delta_{c1}$.

\subsection{Ground-state energy}
To obtain the mean-field ground-state energy $E_G$, we simply need to integrate the gap equation with 
respect to $\Delta$.  This follows because $S(h,q,\Delta)$ is given by  $\frac{\partial E_G}{\partial \Delta} = -2\Delta S(h,q,\Delta)$.
Therefore, up to an overall constant of integration, we have:
\be
\label{Eq:egone}
E_G = - 2\int_0^\Delta dx\, x S(h,q,x) .
\ee
We note that, since the FFLO wavevector $q$ is only defined in the presence of pairing, 
$E_G$ is independent of $q$ in the limit of $\Delta\to 0$.  This can be verified directly 
from examining Eq.~(\ref{eq:eggeneral}) in this limit (it also holds for the free energy $F$, Eq.~(\ref{Eq:mffree})).
This furthermore implies that the abovementioned constant of integration in Eq.~(\ref{Eq:egone}) must be 
$q$ independent, and that differentiating $E_G$ (as approximated by Eq.~(\ref{Eq:egone})) with respect to $q$ leads to the correct equation
for the FFLO wavevector.  

Henceforth we shall assume  $\Delta_{c1}<\Delta<\Delta_{c2}$, and write Eq.~(\ref{Eq:egone}) as:
\be
 E_G =   - 2\int_0^{\Delta_{c1}} dx\, x S(h,q,x) 
 -2 \int_{\Delta_{c1}}^\Delta dx\, x S(h,q,x).
\label{Eq:egsplit}
\ee
The first integral in Eq.~(\ref{Eq:egsplit}) is trivial, because $S$ is $\Delta$-independent in this
 regime.  For the  second integral we'll use our approximate result Eq.~(\ref{closetodeltac}).  Evaluating
the resulting integral finally leads to:
\bea
&&\hspace{-0.75cm}E_G = \!\frac{1}{4\pi}\Delta^2 \ln \frac{q^2}{4\epsilon_b}  \!+ \!\frac{1}{12\pi h}\sqrt{\frac{q_c}{q-q_c}}
  (\Delta \!- \!\Delta_{c1})^2
(2\Delta+ \Delta_{c1}),
\nonumber
\\
&&= \frac{1}{12\pi h} \sqrt{\frac{q_c}{q-q_c}} \big(\Delta_{c1}^3 -\Delta^3 \big) - \Delta^2 S(h,q,\Delta),
\label{Eq:egline1}
\eea
where in the second line we see that $E_G$ does not contain a term linear in $\Delta$ (as one might have
expected from the way we wrote the polynomial in the first line). 

Analytically finding the stationary FFLO wavevector, $q$, turns out to be somewhat tricky despite
the approximations made so far.   To find an approximate formula for $q$ 
 that is valid near the continuous FFLO transition,
it is convenient to differentiate the first line of $E_G$, keeping $\Delta$ arbitrary (i.e.,
not necessarily the stationary solution).  This leads to:
\bea
\label{Eq:dedk}
&&\frac{\partial E_G}{\partial q} = \frac{1}{96\pi h \qb^2(q_c+\qb)}
\\
&&\times
\Big[
6\Delta^2 \qb^{3/2} \big(8h\sqrt{\qb} - \sqrt{2}\sqrt{q_c}(q_c+\qb)(\mu^2-h^2)^{1/4}\big)
\nonumber
\\
&&\nonumber
 -8\Delta^3 \sqrt{q_c\qb}(q_c+\qb)+5\sqrt{2}\sqrt{q_c}\qb^{7/2}(q_c+\qb)(\mu^2-h^2)^{3/4}\Big]
\eea
where we inserted the explicit expression for $\Delta_{c1}$ and we recall that  $\qb \equiv q - q_c$. 
The stable FFLO wavevector satisfies $\frac{\partial E_G}{\partial q} = 0$, but solving this generally
is an arduous task.
 In the vicinity of
the transition, the stationary $\Delta$ and $\qb$ both vanish linearly with $h-\hfflo$.  Thus we shall
Taylor expand the quantity in square brackets in Eq.~(\ref{Eq:dedk}) in small $\Delta$ and $\qb$.   To do this,
we formally replace $\Delta \to \lambda \Delta$ and $\qb\to \lambda \qb$, Taylor expand in small $\lambda$,
and keep terms up to $\curO(\lambda^{3})$.  After setting $\lambda \to 1$  simplifying, we obtain 
\be
\label{Eq:revisedstat}
 0 = \Delta^3 + \frac{3}{2}\Delta^2 \Delta_{c1}- \frac{5}{2} \Delta_{c1}^3 -\frac{6\Delta^2 h}{ q_c^{3/2}}\qb^{3/2},
\ee
the approximate stationarity condition for the FFLO wavevector.  Note that the term that is of highest order
in $\lambda$ (and therefore the smallest) is the final term in Eq.~(\ref{Eq:revisedstat}).  However, 
neglecting this leads to a cubic equation for $\Delta$, with only one real solution that is $\Delta = \Delta_{c1}$.
In terms of $q$, this corresponds to $\qb =  \frac{\sqrt{2}\Delta}{(\mu^2-h^2)^{1/4}}$.  Therefore, within
this approximation the stationary $q$ is adjusted to keep $\Delta$ equal to $\Delta_{c1}$.
However, this solution is unsatisfactory, because we know the correct solution for $\Delta$ is slightly above 
$\Delta_{c1}$, and we are interested in this small difference.

We then proceed by assuming  $\qb = \frac{\sqrt{2}\Delta}{(\mu^2-h^2)^{1/4}}+\epsilon$.  Plugging this into
Eq.~(\ref{Eq:revisedstat}), keeping only terms up to  $\curO(\epsilon)$, solving for $\epsilon$, and 
dropping subdominant terms in $\Delta$, we obtain:
\be
\qb =  \frac{\sqrt{2}\Delta}{(\mu^2-h^2)^{1/4}} - \frac{2^{5/4} h \big(\frac{\Delta}{q_c}\big)^{3/2}}
{(\mu^2-h^2)^{5/8}},
\ee
our final expression for the FFLO wavevector near the FFLO transition.  In the subsequent section, we approximately
solve this simultaneously with the gap equation to study the stable FFLO solution.

\subsection{Solving the remaining equations}
We are now left with the following two equations for $\qb$ and $\Delta$:
\bea
\label{Eq:onesolve}
\qb &=& \frac{\sqrt{2}\Delta}{(\mu^2-h^2)^{1/4}} - \frac{2^{5/4} h \big(\frac{\Delta}{q_c}\big)^{3/2}}
{(\mu^2-h^2)^{5/8}},
\\
\Delta &=& \Delta_{c1}+ h \sqrt{\frac{\qb}{q_c}}\ln \frac{4\epsilon_b}{q^2} ,
\label{Eq:twosolve}
\eea
where $\Delta_{c1}$ is given by Eq.~(\ref{eq:deltac1nearkc}).  We now endeavor to solve these
simultaneously (close to the transition) to establish $\Delta$ and $q$ in the FFLO phase. 
To simplify our analysis, we define 
\bea
\label{Eq:dres}
\Delta  &=& ax + bx^{3/2},
\\
\qb &=& cx + d x^{3/2},
\label{Eq:qhres}
\\
\qt &=& ex + f x^{3/2},
\label{Eq:qtres}
\eea
with $x\equiv \hfflo-h$ is the distance to the transition, which we treat as a small parameter.  
 Recall $\hfflo = \epsilon_b \sqrt{2\mu/\epsilon_b -1}$.
Note
that, although we only obtain a result valid to $\curO(x)$, to obtain this we need to keep
terms up to $\curO(x^{3/2})$.

 Here, $\qt = q_{FFLO}-q>0$, with $q_{FFLO} = 2\sqrt{\epsilon_b}$ the wavevector at the onset of the FFLO phase transition.  
 Clearly, $\qb$ and $\qt$ are
not independent since they both contain $q$ and can be related by Taylor expanding $q_{FFLO}$ in small
$x$.  Doing this leads to the relation
\be
e = \frac{\hfflo}{\sqrt{\epsilon_b} (\mu-\epsilon_b)} -c.
\ee
We now plug Eqs.~(\ref{Eq:dres}), (\ref{Eq:qhres}), and (\ref{Eq:qtres}) into 
Eqs.~(\ref{Eq:onesolve}) and (\ref{Eq:twosolve}), expanding all quantities to $\curO(x^{3/2})$
and demanding equality order by order in $x$.  We find:
\bea
c &=& \frac{\sqrt{2}a}{\sqrt{\mu-\epsilon_b}},
\\
d &=&  \frac{\sqrt{2}b}{\sqrt{\mu-\epsilon_b}} - \frac{a^{3/2} \sqrt{2\mu - \epsilon_b} }{2^{1/4}\epsilon_b^{1/4},
(\mu-\epsilon_b)^{5/4}}
\\
b&=&\frac{1}{\sqrt{2}} d \sqrt{\mu-\epsilon_b} + \frac{1}{\sqrt{2}} e\epsilon_b^{1/4}
 \sqrt{c}\sqrt{\frac{2\mu}{\epsilon_b}-1},
\eea
that lead to the final results 
\bea
a &=& \frac{\sqrt{2}}{3} \frac{\hfflo}{\sqrt{\epsilon_b( \mu-\epsilon_b)}},
\\
e &=& \frac{1}{3}\frac{\hfflo}{\sqrt{\epsilon_b} (\mu-\epsilon_b)}.
\eea
Inserting these results into our equations for $\Delta$ and $q$ leads to our final approximate
results 
\bse
\label{eq:subresults}
\bea
\Delta &\simeq & \frac{\sqrt{2}}{3} \frac{\hfflo}{\sqrt{\epsilon_b( \mu-\epsilon_b)}} (\hfflo-h),
\\
q  &\simeq & 2\sqrt{\epsilon_b}
-\frac{1}{3}\frac{\hfflo}{\sqrt{\epsilon_b} (\mu-\epsilon_b)}(\hfflo-h),
\eea
\ese
showing a continuous onset of FFLO order for $h<\hfflo$.  In Fig.~\ref{fig:two}, we plot these curves
(dashed lines) for the case of $\mu = 3\epsilon_b$.  The red dots show the results obtained by numerically
 minimizing the full mean-field ground state energy, showing agreement close to the transition but
noticeable deviations at lower $h$.

\section{Effect of finite temperature}
We have found that the integral controlling the gap equation for a 2D FFLO state, $S(h,q,\Delta)$  possesses a nonanalytic structure, as a function of
the pairing amplitude $\Delta$, that invalidates a Taylor series in small $\Delta$ (along the lines of Ginzburg-Landau theory)
in the limit of zero temperature.   A 
natural question to ask is whether our findings are relevant given that all experiments
occur at finite temperature.  Although we leave a full analysis of the finite temperature behavior of the 2D FFLO state to future work, 
in this section we briefly touch on this question.

\begin{figure}[ht!]\vspace{-.25cm}
     \begin{center}
        \subfigure{
            \includegraphics[width=85mm]{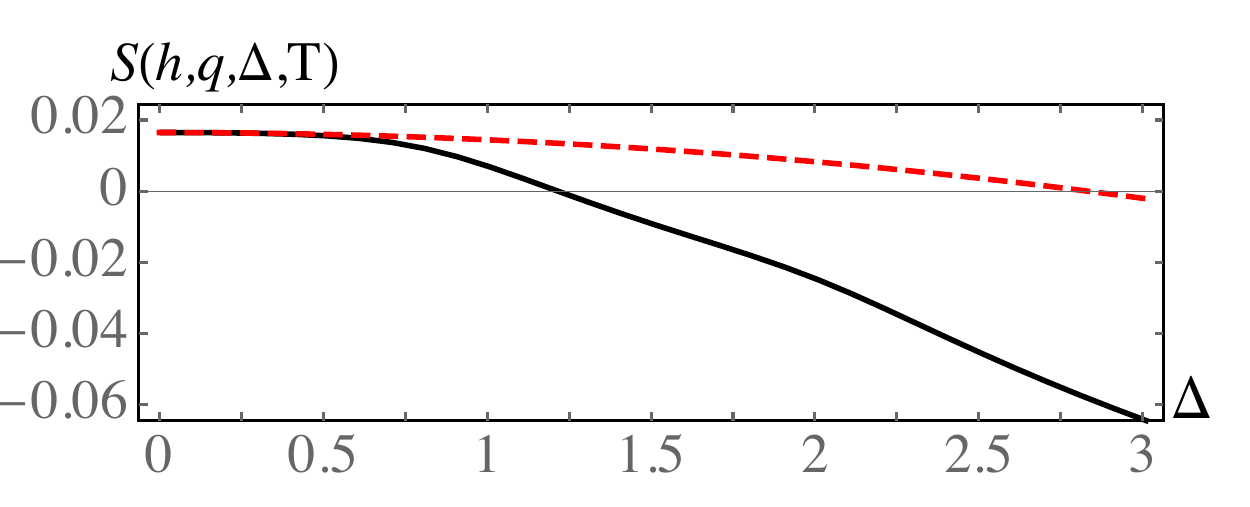}}
        \\ \vspace{-.55cm}
        \subfigure{
            \includegraphics[width=85mm]{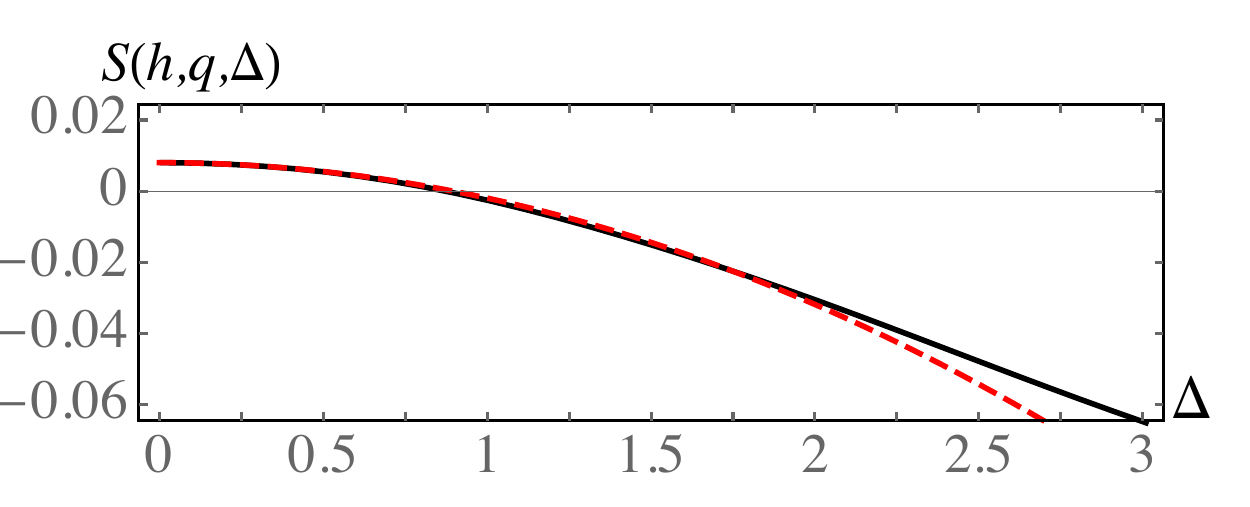}}
    \end{center}\vspace{-.5cm}
    \caption{
        (Color Online) The top panel shows the function $S(h,q,\Delta,T)$  for 
$\mu = 1.3$, $h = 0.5$, and $q=1.8$  (in units such that $\epsilon_b = 1$) and $\kb T= 0.1 \mu$.  The solid black curve is an exact numerical
integration and the red dashed curve is the quadratic order approximation Eq.~(\ref{Eq:essnonzerot}) with 
the coefficients calculated numerically, showing a large discrepancy between the gap equation solutions (given by $S(h,q,\Delta,T) = 0$)
within the two approaches.
The bottom panel shows the same curves but for larger temperature ($\kb T = 0.25\mu$), showing that the quadratic order approximation
holds at larger $T$.
}
   \label{fig:six}
\end{figure}

As depicted in Fig.~\ref{fig:six}, we find that, at finite $T$, 
the singularities in $S(h,q,\Delta)$ are smoothed out.  This can be traced to the
fact that the Fermi functions are no longer sharp steps for $T>0$.  
Therefore,  a Ginzburg-Landau expansion of the form
\be
\label{Eq:essnonzerot}
S(h,q,\Delta,T)  = S(h,q,0,T) + \Delta^2 S'(h,q,0,T) + \cdots ,
\ee
with $S'(h,q,0,T) = \frac{\partial S(h,q,\Delta,T)}{\partial \Delta^2} \Big|_{\Delta^2 \to 0}$, is (technically) valid.
 Here, we have added a temperature argument to this function that is defined in Eq.~(\ref{Eq:fulldelta}).
 However, although such an expansion can be in principle used to solve the gap equation $S(h,q,\Delta,T)=0$,
in practice one typically terminates the expansion at finite $\Delta$, for example at quadratic order as in 
Eq.~(\ref{Eq:essnonzerot}).  Such a truncation fails at sufficiently low $T$, as illustrated in 
Fig.~\ref{fig:six}, which shows the numerically calculated function $S(h,q,\Delta,T)$ for the case of 
$\mu = 1.3$, $h = 0.5$, and $q=1.8$ (in units such that $\epsilon_b = 1$).  In each panel the solid
black curve is the exact numerical integration and the red dashed curve is the quadratic-order
Ginzburg-Landau approximtion.  

  The top panel is for low temperature
($\kb T/\mu = 0.1$), showing that the true gap equation solution (determined by $S(h,q,\Delta,T) = 0$), is at $\Delta \simeq 1.22$.
This is much smaller than that predicted by the quadratic-order Ginzburg-Landau expansion, with the red dashed curve
intersecting zero at $\Delta \simeq 2.83$.     Note that the curves in the top panel do overlap at low 
$\Delta$, showing that the $\Delta$-independent behavior of $S(h,q,\Delta)$ for $\Delta<\Delta_{c1}$ becomes a very small quadratic
order dependence of $S(h,q,\Delta,T)$ at nonzero temperature.  

  The bottom panel compares $S(h,q,\Delta,T)$ to the prediction of the quadratic order approximation to Eq.~(\ref{Eq:essnonzerot})
at a higher $T$, with $\kb T /\mu = 0.25$.  Clearly, these agree well, showing that the Ginzburg-Landau theory does hold at
higher $T$.  Indeed, we must emphasize that, strictly speaking, Ginzburg-Landau theory is valid for $T\to \tc$ and the results
of the present paper do not invalidate this.  However, we expect that our zero temperature results may hold until temperatures
of the order of $\sim 0.1 \mu$.

\section{Concluding remarks}
To conclude, we find that even the simplest 2D FFLO phase, based on a single plane-wave ansatz, has an
 extremely rich structure in which
the gap equation, determining the location of the phase transition and the strength of pairing $\Delta$ in the FFLO 
state possesses nonanalyticities as a function  of $\Delta$ and the FFLO wavevector $q$. 

 We find approximate analytic formulas for the phase boundaries and 
also find  that the equilibrium $\Delta$ can be of the order of the two-body binding energy, making it plausible to 
find the FFLO state in 2D imbalanced Fermi gases.

  Some natural future extensions of this work include determining how the gap equation nonanalyticities 
impact nonzero temperature quantities, applying a similar analysis to more complex FFLO-type phases 
(e.g., of the Larkin-Ovchinnikov type with $\Delta(\br) = \Delta \cos \bq\cdot \br$) and analyzing
 analyzing fluctuation effects in 2D~\cite{Leo2011}.

\section*{Acknowledgements}
We gratefully acknowledge useful discussions with Leo Radzihovsky, Stephen Shipman, and Ilya Vekhter.
This work was supported by the National Science Foundation Grant No. DMR-1151717.
This work was supported in part by the National Science Foundation under 
Grant No. PHYS-1066293 and the hospitality of the Aspen Center for Physics.
We also acknowledge support from the
German Academic Exchange Service (DAAD) and the hospitality of the 
Institute for Theoretical Condensed Matter Physics at the Karlsruhe Institute of 
Technology.

\appendix

\section{Quartic equation}
\label{sec:appquar}
In this section we recall the solution to the quartic equation~\cite{Abramowitz}.  A general quartic equation is of the form
\be
\label{eq:quarticequation}
ax^4 + bx^3 +cx^2 +dx +e = 0,
\ee
which, comparing to Eq.~(\ref{Eq:fullquartic}), we'll need for 
 $a=1$, $b=0$ and 
\bea
c&=& - (4\mut+q^2\cos^2\theta),
\\
d&=& -4hq\cos\theta,
\\
e&=&  4(\Delta^2-h^2+\mut^2) .
\eea
The general solution to the quartic equation is~\cite{Abramowitz}: 
\bse
\label{Eq:generalquartic}
\bea
\label{Eq:x12}
x_{1,2} &=& - \frac{b}{4a} - \Sh \pm \frac{1}{2}\sqrt{-4\Sh^2 -2p+\frac{\qh}{\Sh}}.
\\
x_{3,4} &=&   - \frac{b}{4a} + \Sh \pm \frac{1}{2}\sqrt{-4\Sh^2 -2p-\frac{\qh}{\Sh}},
\label{Eq:x34}
\eea
\ese
where $x_2$ and $x_4$ take the $-$ in each line.

Here we defined:
\bea
\label{Eq:peeee}
p&=& \frac{8ac-3b^2}{8a^2} = -4\mut -q^2 \cos^2 \theta,
\\
\qh &=& \frac{b^3 - 4abc + 8a^2 d}{8a^3} = -4hq\cos \theta ,
\eea
where the final equalities apply to the present case.

We also define:
\bea
\label{Eq:bigess}
\Sh &=& \frac{1}{2} \sqrt{-\frac{2}{3}p + \frac{1}{3a} \big( Q + \frac{\Delta_0}{Q}\big)},
\\
Q &=& \sqrt[3]{\frac{\Delta_1 + \sqrt{\Delta_1^2 - 4\Delta_0^3}}{2}},
\label{Eq:bigque}
\eea
%
Here,  $\Delta_0$ and $\Delta_1$ are :
\bea
\Delta_0 &=& c^2 -3bd +12ae  ,
\\
 &=& 48(\Delta^2-h^2 +\mut^2) + (-4\mut - q^2 \cos^2\theta)^2,
\eea
and 
\bea
&& \Delta_1 = 2c^3 - 9bcd + 27 b^2 e + 27 ad^2 - 72 ace ,
\\
&&= 432 h^2 q^2 \cos^2 \theta + 288(\Delta^2 - h^2 +\mut^2)(4\mut + q^2\cos^2 \theta )
\nonumber 
\\
&& 
-2(4\mut + q^2 \cos^2 \theta)^3.\label{eq:deltaone}
\eea
Note that $\Sh$ and $Q$ are too complex to write out explicitly for the present case.  
In terms of $\Delta_0$ and $\Delta_1$, the discriminant $\Delta_d$ is
\be
\label{eq:discriminant}
\Delta_d = - \frac{1}{27}\big(\Delta_1^2 - 4\Delta_0^3\big).
\ee
Additionally, $\Delta_d$ can be written in terms of the quartic equation solutions as 
\be
\label{Eq:discsols}
\Delta_d = \sum_{i<j} (x_i - x_j)^2,
\ee
with the sum being over all pairs of solutions.

The quartic equation solutions are further characterized by the
functions
\bea
\label{Eq:bigpee}
P &=& 8ac - 3b^2 = -8(4\mut +q^2\cos^2 \theta),
\\
D &=& 
64a^3 e - 16a^2c^2 + 16ab^2 c - 16 a^2 bd - 3b^4 
\\
&=& 
256(\Delta^2 -h^2+\mut^2) -16(4\mut+q^2\cos^2\theta)^2,
\eea
which determine the nature of the roots in various regimes.  
Our $p_1$ and $p_2$ are given by $x_{3}$ and $x_4$ since $x_{1,2}$ are negative for 
our parameters (when they are real). 
Therefore we define:
\bse
\label{Eq:peesAPP}
\bea
 p_2(\theta) &=& S + \frac{1}{2}\sqrt{-4S^2 -2p-\frac{\qh}{S}},
\\
p_1(\theta) &=& S - \frac{1}{2}\sqrt{-4S^2 -2p-\frac{\qh}{S}}.
\eea
\ese
%
For angles $\theta$ such
that $p_1$ and $p_2$ are real,  these will define the boundaries of the regions where $E_{\bp\pm}<0$.
However, in the main text we'll need these solutions even when they are complex, which occurs for angle $\theta$ that 
do not intersect (for any radial $p$) a region where $E_{\bp\pm}<0$.

\section{Approximate calculation of $S_+$}
\label{sec:appsp}
In this section, we approximately evaluate the integral $S_+$ in the region $\Delta\alt \Delta_{c1}$.
A direct integration of the radial $p$ integral gives 
\be
\label{Eq:essplusexpr}
S_+ = \frac{1}{4\pi^2}\int_{\theta_c}^\pi d\theta \, \ln \frac{p_1^2-2\mut +2E_{p_1}}{p_2^2-2\mut +2E_{p_2}},
\ee
where $\theta_c$ is the point where $p_1$ and $p_2$ cross,
with $p_{1,2} = S \pm \frac{1}{2}\sqrt{-4S^2 -2p-\frac{\qh}{S}}$ as previously.
  To find approximate results for $p_1$ and $p_2$
we define small parameters $\delta = \Delta- \Delta_{c1}$ and $\qb = q-q_c$, and approximate
$\Delta_{c1}$ by its form near $q_c$, given in Eq.~(\ref{eq:deltac1nearkc}).  This implies
\be
\delta = \Delta- \frac{1}{\sqrt{2}} (\mu^2 - h^2)^{1/4}\qb.
\label{eq:littledeltadef}
\ee
For $\Delta\alt \Delta_{c1}$, the angle integral of $S_+$ is restricted to the immediate vicinity
of $\theta = \pi$, allowing us to simultaneously expand in the small parameters 
$y = \cos\theta + 1$, $\qb$, and $\delta$. 

Within this approximation, we find, for the following quantities entering
$p_{1,2}$, 
\bea
&&\Sh \simeq  \frac{1}{4}h\Big( \frac{\Delta^2 +8(\mu^2-h^2)}{q_c(\mu^2-h^2)} 
-\frac{(2\qb+q_c)y}{\sqrt{\mu^2-h^2}}\Big),
\label{Eq:finals}
\\
&&-4\Sh^2 -2p-\frac{\qh}{\Sh}   \simeq \frac{4h^2\qb(y_c-y)}{q_c\sqrt{\mu^2-h^2}},
\eea
where $y_c= 1+\cos\theta_c$ is given by
\be
y_c \simeq - \frac{\delta q_c}{2\sqrt{2}h^2\qb } \big( \sqrt{2}\delta + 2\qb (\mu^2 - h^2)^{1/4}\big) .
\ee
After using the preceding results in Eq.~(\ref{Eq:essplusexpr}), we
Taylor expand to leading order in the small parameters $\delta$, $\qb$ and $y$, and take the limit of small $y-y_c$, 
arriving at
\be
S_+ = \frac{1}{4\pi^2} \int_{\theta_c}^\pi d\theta \, \frac{-8\sqrt{2}h(\mu^2-h^2)^{3/4} \sqrt{\frac{\qb}{q_c}} }{a-2by} \sqrt{y_c-y},
\ee
where we defined 
\bea
a &=& 2\sqrt{2}\qb(\mu^2-h^2),
\\
b&=& h^2\Big( \frac{2\sqrt{2}\mu}{q_c} -\frac{q_c}{\sqrt{2}}\Big) = \frac{2\sqrt{2}h^2\sqrt{\mu^2-h^2}}{q_c},
\eea
which, in the parameter range of interest, are always positive.  

The range of the integral is the vicinity of $\pi$.  In this limit, we can write
\be
y \simeq  1+\cos\theta \approx \frac{1}{2}(\pi-\theta)^2.
\ee
Then, defining  $ x= \pi -\theta$, we have $y =\frac{1}{2}x^2$ and $S_+$ is :
\be
S_+ \simeq -\frac{2}{\pi^2} \sqrt{\frac{\qb}{q_c}}h(\mu^2-h^2)^{3/4} \int_0^{x_c} dx \frac{\sqrt{x_c^2-x^2}}{a-bx^2},
\ee
where we defined $x_c^2 = 2y_c$.
We can evaluate the $x$ integral, assuming the denominator does not vanish:
\be
\int_0^{x_c} dx \frac{\sqrt{x_c^2-x^2}}{a-bx^2} = \frac{\pi}{2b} \big( 1-\frac{1}{\sqrt{a}}\sqrt{a-bx_c^2}\Big).
\ee
After plugging in and substituting the definition of $\delta$ in terms of $\Delta$, we obtain 
\bea
S_+ 
&=&- \frac{\sqrt{\qb q_c}(\mu^2-h^2)^{1/4}}{2\sqrt{2}\pi h}
\Big( 1 - \sqrt{\frac{1}{2} +\frac{\Delta^2}{2\Delta_{c1}^2}}\Big),
\eea
where we used the definition of $\Delta_{c1}$ to this order, i.e.,  Eq.~(\ref{eq:deltac1nearkc}). 
This expression is only valid to leading order in the simultaneous small parameters $\qb$ and 
$\Delta- \Delta_{c1}$.  The quantity in parentheses vanishes for $\Delta\to \Delta_{c1}$; to leading order
in small $\Delta- \Delta_{c1}$ the quantity in parentheses is :
\be
\Big( 1 - \sqrt{\frac{1}{2} +\frac{\Delta^2}{2\Delta_{c1}^2}}\Big) \simeq\frac{1}{2}\big(1-\frac{\Delta}{\Delta_{c1}}\big).
\ee
Using this, and again using Eq.~(\ref{eq:deltac1nearkc}), we arrive at  Eq.~(\ref{Eq:spluslast}) from the main text.


\begin{thebibliography}{10}
\bibitem{Balancednote}
Here and elsewhere, the term \lq\lq balanced\rq\rq\ means the equilibrium numbers are the same, $N_\uparrow = N_\downarrow$, while
imbalanced means the equilibrium numbers are different ($N_\uparrow>N_\downarrow$).
\end{thebibliography}
\end{document}